\documentclass[11pt]{article}
\usepackage{amsmath, graphicx, fullpage, parskip}
\usepackage{natbib}
\usepackage{url}

\usepackage{amssymb, amsthm}
\usepackage{algorithm}
\usepackage{algpseudocode}
\usepackage{amsmath, amssymb, amsthm}
\usepackage{float,multirow,subcaption, setspace}
\usepackage{graphicx}
\usepackage{comment}
\usepackage{enumitem}
\usepackage{tikz}
\usepackage{bm, bbm}
\usetikzlibrary{shapes.geometric, positioning}
\usetikzlibrary{quotes, angles}
\usepackage{rotating}
\usepackage{hyperref}

\definecolor{SkyBlue}{RGB}{14, 118, 188}
\definecolor{BrightRed}{RGB}{223, 82, 78}
\definecolor{Green638}{RGB}{165,255,118} 
\definecolor{RevisedColor}{RGB}{25, 200, 150}

\hypersetup{pdfborder = {0 0 0.5 [3 3]}, colorlinks = true, linkcolor = BrightRed, citecolor = SkyBlue}

\newcommand{\skd}[1]{\textcolor{cyan}{\small [skd]: #1}}

\newcommand{\R}{\mathbb{R}} 
\def\P{\mathbb{P}} 

\newcommand{\calG}{\mathcal{G}}
\newcommand{\calX}{\mathcal{X}}
\newcommand{\calC}{\mathcal{C}}
\newcommand{\calA}{\mathcal{A}}
\newcommand{\calD}{\mathcal{D}}
\newcommand{\calK}{\mathcal{K}}
\newcommand{\calE}{\mathcal{E}}

\newcommand{\cutset}{\calC}

\newcommand{\ind}[1]{\mathbbm{1}\left( #1 \right)} 

\newcommand{\normaldist}[2]{\mathcal{N}\left(#1,#2\right)} 
\newcommand{\unifdist}[2]{\textrm{Uniform}\left(#1,#2\right)} 

\newcommand{\by}{\bm{y}}
\newcommand{\bx}{\bm{x}}

\newcommand{\br}{\bm{r}}

\newcommand{\pcont}{p_{\text{cont}}}
\newcommand{\pcat}{p_{\text{cat}}}

\newcommand{\nx}{\texttt{nx}}
\newcommand{\nxl}{\texttt{nxl}}
\newcommand{\nxr}{\texttt{nxr}}

\newcommand{\bmu}{\boldsymbol{\mu}}

\theoremstyle{plain}
\newtheorem{theorem}{Theorem}

\theoremstyle{definition}

\theoremstyle{plain}
\newtheorem{remark}[theorem]{Remark}

\newcommand{\revised}[1]{#1}
\title{\textbf{flexBART}: Flexible Bayesian regression trees with categorical predictors}
\author{Sameer K. Deshpande\thanks{Dept.~of Statistics, University of Wisconsin--Madison. \texttt{sameer.deshpande@wisc.edu}}}

\begin{document}
\maketitle
Most implementations of Bayesian additive regression trees (BART) one-hot encode categorical predictors, replacing each one with several binary indicators, one for every level or category.
Regression trees built with these indicators partition the discrete set of categorical levels by repeatedly removing one level at a time.
Unfortunately, the vast majority of partitions cannot be built with this strategy, severely limiting BART's ability to partially pool data across groups of levels.
Motivated by analyses of baseball data and neighborhood-level crime dynamics, we overcame this limitation by re-implementing BART with regression trees that can assign multiple levels to both branches of a decision tree node.
To model spatial data aggregated into small regions, we further proposed a new decision rule prior that creates spatially contiguous regions by deleting a random edge from a random spanning tree of a suitably defined network. 
Our re-implementation, which is available in the \textbf{flexBART} package, often yields improved out-of-sample predictive performance and scales better to larger datasets than existing implementations of BART.

\newpage

\section{Introduction}
\label{sec:introduction}
\revised{The Bayesian additive regression trees \citep[BART;][]{Chipman2010} model approximates unknown functions using ensembles of regression trees.}
When \revised{the target function} is non-linear or involves complicated high-order interactions between several predictors, correctly specifying \revised{its} functional form with a parametric model is extremely challenging, if not impossible.
With BART, however, users can \revised{accurately predict evaluations of a function and obtain reasonably well-calibrated uncertainties \textit{without pre-specifying the functional form of the function.}}
BART's ease-of-use and generally excellent performance have made it a popular ``off-the-shelf'' modeling tool.

\revised{Most BART implementations \emph{one-hot encode} categorical predictors, replacing each with several binary indicators, one for each level or category.}
We demonstrate that one-hot encoding severely limits BART's ability to ``borrow strength'' across different levels of a categorical variable.
We re-implemented BART in a way that permits much more flexible modeling with categorical predictors \revised{and generally yields improved predictions.
The following problems motivate our work.}

\textbf{Pitch framing in baseball.} \revised{In baseball, when batters do not swing at pitches, umpires classify the pitches as balls or strikes.
Ball/strike decisions are quite noisy and depend not only on pitch location but also on the umpires and the players involved.
\citet{DeshpandeWyner2017} estimated each umpire's called strike probabilities using a hierarchical logistic regression model that assumed player effects were constant across umpire and location.
Despite this arguably over-simplified assumption, their model displayed reasonably promising predictive results, misclassifying about 10\% of umpire decisions. 
Can a more sophisticated model, which allows some (but not all) catchers to influence some (but not all) umpires at some (but not all) locations, provide more accurate predictions?}

\begin{figure}[ht!]
\centering
\begin{subfigure}[t]{0.45\textwidth}
\centering
\includegraphics[width = \textwidth]{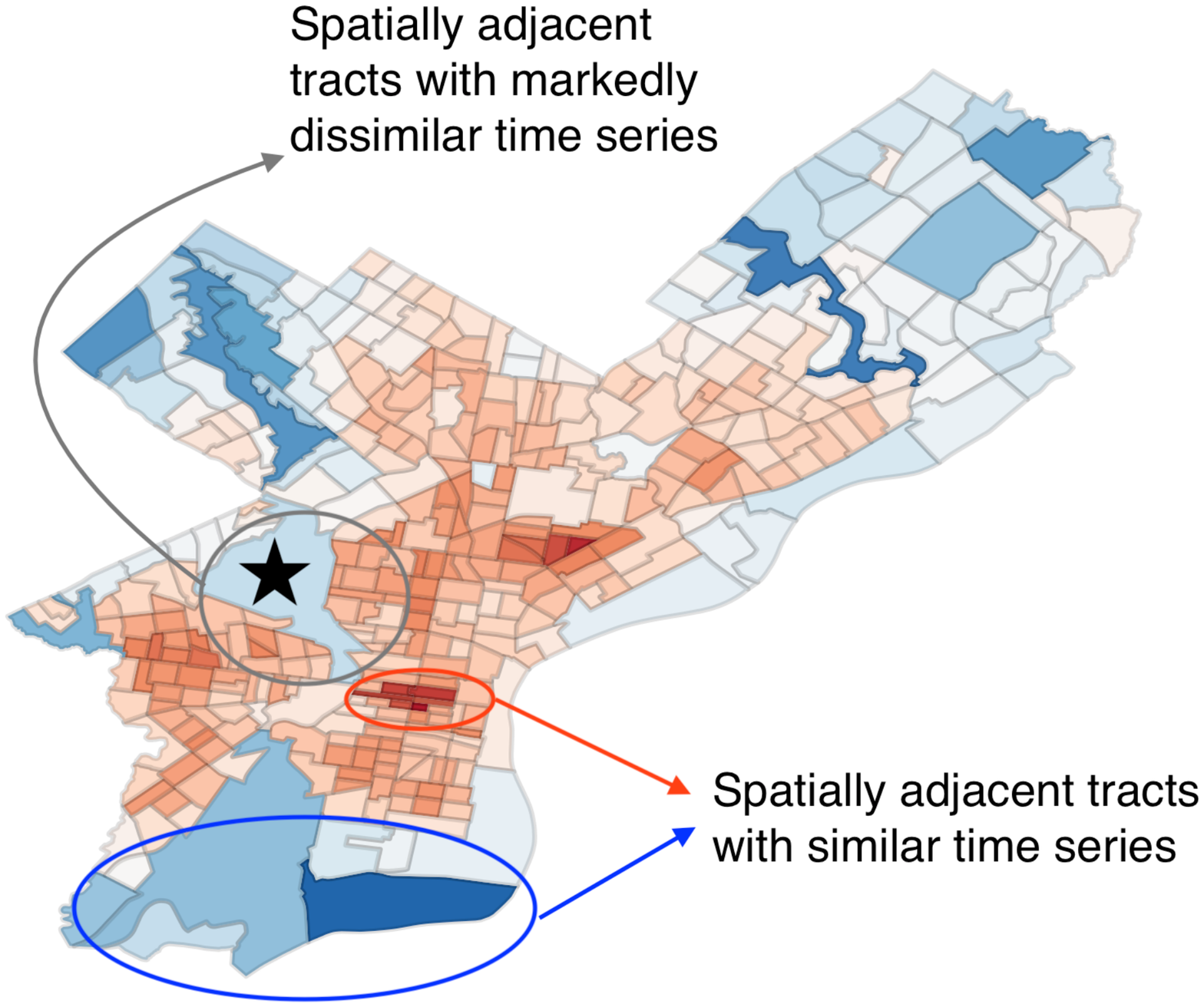}
\caption{}
\label{fig:philly_means}
\end{subfigure}
\begin{subfigure}[t]{0.45\textwidth}
\centering
\includegraphics[width = \textwidth]{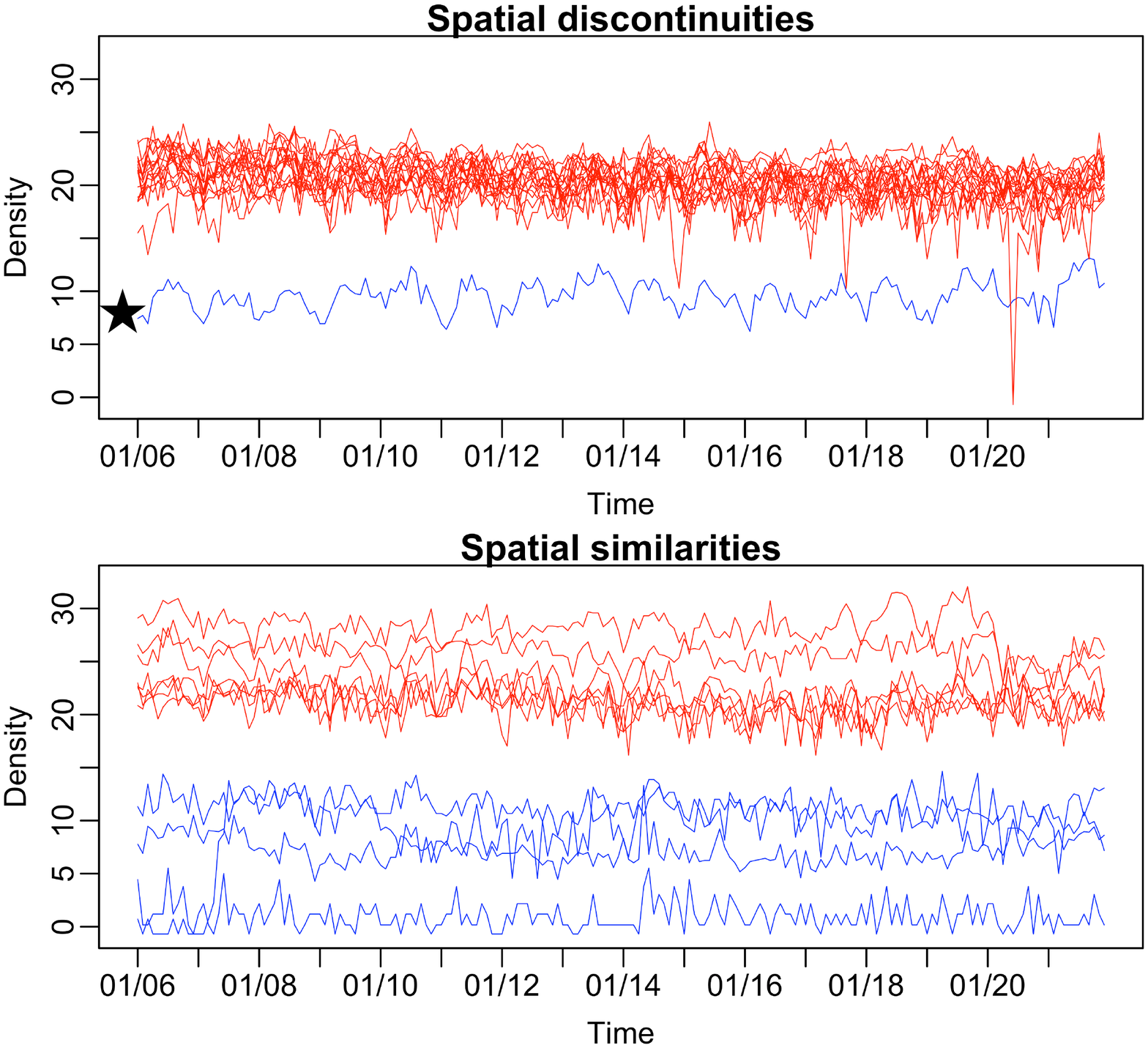}
\caption{}
\label{fig:philly_combined}
\end{subfigure}
\caption{(a) Average crime densities in each census tracts, with red (resp.\ blue) corresponding to higher (resp.\ lower) crime densities. (b) Time series of the circled census tracts in (a) displaying sharp spatial discontinuities (top) and spatial similarity (bottom). The starred tract in (a) corresponds to the lowest curve in the top panel of (b).}
\label{fig:philly_crime_intro}
\end{figure}

\revised{To investigate this possibility, it is tempting to elaborate \citet{DeshpandeWyner2017}'s model with two- and three-way interactions between umpires, pitch location, and players.
Unfortunately, including such interactions quickly exhausts available degrees of freedom: in each season, there are about 100 umpires, 1000 batters, 100 catchers, and 700 pitchers and only 350,000 called pitches.
Tree-based models, which can more parsimoniously accommodate complicated higher-order interactions, are a potentially more promising alternative.}
\textbf{Crime in Philadelphia}. \revised{Figure~\ref{fig:philly_means} shows the mean monthly crime density\footnote{Defined as crime counts per square mile} in each census tract in the city of Philadelphia averaged over the 16 year period from January 2006 to December 2021. 
Although there are some spatially adjacent tracts that have very similar averages and temporal trends (Figure~\ref{fig:philly_combined}), other tracts display markedly different pattens than their neighbors (e.g., the starred tract and top panel of Figure~\ref{fig:philly_combined}).
Standard spatial smoothing techniques like conditional autoregressive priors can over-smooth across these discontinuities, producing highly biased crime forecasts \citep{Balocchi2019}.}
\revised{An ideal model of the spatiotemporal variability in crime density must (i) account for non-linearities in time; (ii) allow those non-linearities to vary across tracts; and (iii) exploit spatial similarities when they are present while adapting to discontinuities suggested by the data.}
Unfortunately, correctly pre-specifying such non-linear interactions in a parametric model is challenging, motivating a more flexible, nonparametric approach. 

\subsection{Our contributions}

In both problems, we must estimate a function that may (i) be highly non-linear; (ii) depend on the complex interaction and interplay between categorical and continuous covariates; and (iii) have a difficult-to-specify functional form.
From this perspective, BART appears ideally suited for both problems.
As we argue conceptually and demonstrate empirically, however, existing implementations of BART can be improved with a more thoughtful treatment of the categorical predictors.
Briefly, in the baseball application, by one-hot encoding player and umpire identities, existing implementations of BART are unable to pool data across many small groups of players and umpires.
\revised{And} in the crime application, these implementations cannot adapt to any pockets of spatial smoothness.

\revised{We re-implemented BART with trees that assign multiple levels of a categorical predictor to both the left and right branches of a decision node.
Our new implementation explores a much richer space of partitions of the categorical levels and often yields better predictions compared to default implementations of BART that use one-hot encoding.
We further introduce several new decision rule priors that allow trees to partition networks recursively into connected components, allowing us to incorporate spatial adjacency information directly in our BART models.}

The remainder of the paper is structured as follows.
\revised{
We review the basic BART model in Section~\ref{sec:bart_review}.
In Section~\ref{sec:partition_problem}, we first reveal how one-hot encoding} severely restricts the representational flexibility of regression tree ensembles.
\revised{We then describe our implementation and introduce a new decision rule prior that allows us to fit BART models over networks.}
We illustrate the improvements in predictive quality and runtime offered by our implementation \revised{on both synthetic data (Section~\ref{sec:simulations}) and real-world data (Section~\ref{sec:real_data}).}
We conclude in Section~\ref{sec:discussion} with a discussion of several methodological and computational extensions.

\section{A brief review of BART}
\label{sec:bart_review}
In each motivating application, we want to estimate a function without imposing strong structural assumptions about its form.
BART facilitates such estimation by approximating the function with a sum of binary regression trees.
For simplicity, we describe BART only in the homoscedastic setting with independent Gaussian errors.

Suppose that we observe $n$ pairs $(\bx_{1}, y_{1}), \ldots, (\bx_{n}, y_{n})$ of predictors $\bx \in \calX$ and scalar outcomes $y \in \R$ from the model $y \sim \normaldist{f(\bx)}{\sigma^{2}}.$
We assume throughout that we have $\pcont$ continuous predictors and $\pcat$ categorical predictors, for a total of $p = \pcont + \pcat$ predictors.
Without loss of generality, we assume that the continuous predictors have been re-scaled to lie in the interval $[0,1].$
For $j = 1, \ldots, \pcat,$ we assume that the $j$-th categorical predictor takes values in a discrete set $\mathcal{K}_{j}.$
We concatenate our predictors into a single vector $\bx = (\bx^{\top}_{\text{cont}}, \bx^{\top}_{\text{cat}})^{\top}$ whose first $\pcont$ entries record the values of the continuous predictors and whose last $\pcat$ entries record the values of the categorical predictors.
This way, $\calX$ is a product space: $\calX= [0,1]^{\pcont} \times \mathcal{K}_{1} \times \cdots \times \mathcal{K}_{\pcat}.$

Formally, a \textit{regression tree} over $\calX$ is a triplet $(T, \calD, \bmu)$ consisting of (i) a finite, rooted binary tree $T$ containing several terminal or \textit{leaf nodes} and non-terminal nodes; (ii) a collection of decision rules $\calD,$ one for each non-terminal node in $T;$ and (iii) a collection $\bmu$ of scalars or \textit{jumps}, one for each leaf node in $T.$
Every non-terminal node is connected to two children nodes, a left child and a right child.
Decision rules take the form $\{X_{j} \in \cutset\}$ where $\cutset$ is a subset of the available values of the $j$-th covariate $X_{j}$ at the associated non-terminal node of $T.$
When $X_{j}$ is continuous, $\cutset$ is a half-open interval $[0,c) \subset [0,1]$ and when $X_{j}$ is categorical, $\cutset$ is a proper, non-empty subset of $\calK_{j - \pcont}.$ 

Given a \textit{decision tree} $(T,\calD)$ and an $\bx \in \mathcal{X},$ we can trace a path \revised{from the root to a unique leaf} by following the decision rules.
Starting from the root node, when the path reaches a node with decision rule $\{X_{j} \in \cutset\},$ it proceeds to the left child if $x_{j} \in \cutset$ and to the right child otherwise.
\revised{In this way, $(T, \calD)$ partitions $\mathcal{X}$ into several regions, one for each leaf.}
By associating each leaf with its own scalar, the regression tree $(T, \calD, \bmu)$ represents a piecewise constant function over $\calX.$

In principle, we can always find a piecewise constant function that approximates a function $f$ arbitrarily well.
When $f$ involves non-linearities or complicated interactions, the approximating function typically corresponds to a very deep regression tree with many leaf nodes.
Rather than attempt to learn a single, deep tree, BART instead learns an ensemble of much simpler trees that sum to an accurate approximation of $f.$

\subsection{Drawing regression trees from the prior}
\label{sec:cgm98_prior}

\revised{
\citet{Chipman1998} introduced a regression tree prior that strongly regularizes towards relatively shallow trees with only a few leaves.}
We describe the regression tree prior implicitly with a three-step procedure for simulating prior draws.
We first draw $T$ by simulating a branching process.
Then, conditionally on $T,$ we sequentially draw random decision rules at each non-terminal node.
Finally, we draw the $L(T)$ jumps in $\bmu.$ 

\textbf{Drawing $T$}. We grow $T$ recursively starting from a single root node, which is considered to be a terminal node of depth zero.
Each time a new terminal node is created at depth $d$, we grow the tree from that node by creating two child nodes with probability $0.95(1 + d)^{-2}.$
Because the growth probability decays quadratically with tree depth, the branching process almost surely terminates with finitely many leaves.

\textbf{Drawing $\calD \vert T.$} Drawing independent decision rules can produce trees that do not partition $\calX.$
We instead draw decision rules sequentially, only sampling the rule at a non-terminal node after  all its ancestors in $T$ have an associated rule.
We draw the rule at a given non-terminal node by first picking $j,$ the index of the decision variable, uniformly at random.
Then, we set $\cutset$ to be a random subset of $\calA,$ the set of all available values of $X_{j}$ at the current tree node.
We can compute $\calA$ by taking intersections of the $\calC$'s (or their complements) at the node's ancestors in the tree. 
When $X_{j}$ is continuous, $\calA$ is an interval and we set $\cutset = [0,c)$ where $c$ is drawn uniformly from $\calA.$ 
When $X_{j}$ is categorical, $\calA$ is a discrete set and we simply assign each element of $\calA$ to $\cutset$ with probability $1/2.$

\textbf{Drawing $\bmu \vert T, \calD.$} We draw the scalar jump for each leaf independently from a $\normaldist{\mu_{0}}{\tau^{2}/M}$ distribution.
In this way, the marginal prior of $f(\bx)$ is $\normaldist{\mu_{0}}{\tau^{2}}$ for every $\bx \in \calX.$
\citet{Chipman2010} recommended setting $\mu_{0}$ and $\tau$ so that this marginal distribution places substantial probability over the range of the observed $y_{i}$'s.

\subsection{Posterior computation}
\label{sec:posterior_computation}

\citet{Chipman2010} used a Gibbs sampler to simulate posterior samples of the regression tree ensemble.
In each iteration, every regression tree is updated conditionally fixing the remaining $M-1$ trees and $\sigma.$
Then, $\sigma$ is updated conditionally fixing all regression trees.
Two steps are used to update an individual regression tree $(T, \calD, \bmu).$
First, given the data, $\sigma,$ and all other trees, a new decision tree $(T,\calD)$ is drawn from its marginal posterior.
Then, a new collection of jumps $\bmu$ is drawn conditionally on the just-drawn $(T, \calD).$

\revised{The new decision tree $(T, \calD)$ is drawn with a Metropolis-Hastings step in which we randomly propose pruning or growing the tree.}
Pruning $(T,\calD)$ involves (i) deleting two leafs sharing a common parent and their incident edges in $T$; (ii) deleting the decision rule associated with the parent of the deleted leafs; and (iii) leaving the rest of the tree and decision rules unchanged.
Growing $(T,\calD)$ in contrast, involves (i) selecting an existing leaf and connecting two new children to it; (ii) drawing a new decision rule for the selected node; and (iii) leaving the rest of the tree and decision rules unchanged.
When growing a tree, we draw the new decision rule from the prior to simplify the acceptance probability calculation; see Appendix~\ref{app:calculations}.

In our simple regression setting with i.i.d.\ errors, the jumps in $\bmu$ are conditionally independent and normally distributed.
When the number of observations in leaf is large, the conditional posterior distribution of corresponding jump is sharply concentrated around the average of a particular partial residual (see Equation~\eqref{eq:mu_posterior}).
In this way, each tree performs adaptive partial pooling by partitioning observations into several cells and then averaging within each cell.

\revised{During grow moves, we propose the new decision rule from the prior.
One might expect that a more informative, data-based proposal distribution might lead to more efficient MCMC exploration of tree space.
As we argue in Appendix B, however, informative proposals can be counter-productive.
In grow moves, the acceptance probability depends on the ratio of prior and proposal densities.
When the proposal is highly informative, this ratio is much smaller than one, deflating the acceptance probability and inhibiting the Markov chain's ability to explore tree space.
\citet{Zhou2022} observed a similar phenomenon in Bayesian variable selection; see Example 1 of that paper.}

\section{Introducing flexBART}
\label{sec:flex_bart}
\subsection{A one-hot encoded hole in the BART prior}
\label{sec:partition_problem}
Most BART implementations one-hot encode categorical predictors, replacing each with a collection of binary indicators, one for every level.
That is, rather than work with the original predictor space $\calX,$ which is a product of Euclidean and discrete spaces, they instead work with an extended predictor space $\tilde{\calX} = [0,1]^{\tilde{p}}$ where $\tilde{p} = \pcont + \lvert \calK_{1}\rvert + \cdots + \lvert \calK_{\pcat}\rvert.$
At first glance, one-hot encoding appears innocuous since no information about the predictor is lost.
\revised{In fact,} although the standard BART prior induces a prior over a vast class of partitions of the extended, continuous space $\tilde{\calX},$ it places zero probability on a staggeringly overwhelming majority of partitions of the original space $\calX.$

\revised{To see this, consider a setting with only one categorical predictor $X$ taking values in $\{c_{1}, \ldots, c_{K}\}.$
One-hot encoding replaces $X$ with $K$ binary indicators, $X_{1}, \ldots, X_{K},$ where $X_{k} = \ind{X = c_{k}}.$ }
Suppose that $(T, \calD, \bmu)$ is a regression tree with at least two leaf nodes defined over the extended space $[0,1]^{K}.$
\revised{Through its partition of $[0,1]^{K},$ the tree $(T, \calD)$ partitions the points $\{\bm{e}_{1}, \ldots, \bm{e}_{K}\},$ where the $k$-th entry of $\bm{e}_{k}$ is equal to one and all other entries are zero.
The partitions of the $\bm{e}_{k}$'s in turn induces a partition over $X$'s levels, as we can associate each point $\bm{e}_{k}$ with level $c_{k}.$}

\revised{Since the decision rule $\{X_{j} < c\}$ sends only $\bm{e}_{j}$ to the right and all other $\bm{e}_{j'}$'s to the left, $(T, \calD)$ can only form three types of partitions of $X$'s levels: (i) the partition placing all levels in one set; (ii) the partition placing each level in its own singleton set; and (iii) a partition with $1 \leq S \leq K-2$ singletons and one set containing $K-S$ elements.
There are $\binom{K}{K-S}$ such partitions of the last type and so the total number of partitions that can be formed after one-hot encoding is $2^{K}-K.$}
When $K > 2,$ this number is much smaller than the $K$-th Bell number, which counts the number of partitions of $K$ objects \revised{and is lower bounded by $(K/\log_{2}(K))^{K}$ for large $K$ \citep{BerendTassa2010}.}
For instance, after one-hot encoding, the BART prior places positive probability on just 27 of the 52 possible partitions when $K = 5$ and on less than 1\% of partitions when $K = 10$ (1,014 out of 115,975).

\revised{To encourage more diverse partitions of categorical levels, we forgo one-hot encoding and implement BART using the prior described in Section~\ref{sec:cgm98_prior}.
Although the decision rule prior that independently assigns levels to the left and right might be reasonable for the motivating baseball example, it is not wholly satisfactory for our crime data as it ignores the spatial adjacency and tends not to produce spatially contiguous clusters of census tracts (see Figure~\ref{fig:philly_partitions}).}
\subsection{Splitting a network-structured categorical predictor}
\label{sec:network_splits}
\revised{To form more structurally-informed partitions of categorical levels, we must modify the decision rule prior.
Although we are motivated primarily by spatial applications, our modification is somewhat more general.
Specifically, suppose the levels of a categorical $X$ display a known similarity structure and that we wish for similar levels to be clustered together more often than not.}
To achieve this, we form a network $\calG$ \revised{whose vertices are the levels of $X$ and whose edges are drawn between similar levels.} 
\revised{Now suppose that in drawing a regression tree the prior that we have selected $X$ as the decision variable.}
Further, suppose that the set of available levels $\calA$ induces a connected subgraph $\calG[\calA]$ of $\calG.$ 
Unless $\calG[\calA]$ is a complete graph, independently assigning levels in $\calA$ to $\calC$ with a fixed probability tends not to partition $\calG[\calA]$ into two connected components.
\revised{It suffices, therefore, to specify a process for partitioning the connected subgraph $\calG[\calA]$ into two connected components.}

\revised{To this end, we propose four different network splitting strategies, which are based on two general approaches for partitioning fully connected networks.
The first, which is popular in the network analysis literature \citep[see, e.g.,][]{Li2020_hierarchical} is based on \emph{Fiedler vector}, which is eigenvector associated with the smallest non-zero eigenvalue of the network's Laplacian matrix\footnote{Given a network with adjacency matrix $A,$ the Laplacian is defined as $L = D - A,$ where $D$ is a diagonal matrix containing the row sums of $A$}.
The vertices corresponding to positive (resp.\ negative) entries of the Fiedler vector form a connected component of the original network \citep{Fiedler1973}. 
Another approach, which is popular in the spatial clustering literature \citep[see, e.g.,][]{Teixeira2015, Teixeira2019, LiSang2019, Luo2021}, partitions a network into connected components by removing edges from randomly drawn spanning trees of the network; see Figure~\ref{fig:mst_split}.}

\revised{Our first strategy (\texttt{gs1}) computes the deterministic Fielder partition of $\calG[\calA].$
The remaining strategies begin by first drawing a spanning tree of $\calG[\calA]$ uniformly at random using Wilson's algorithm \citep{Wilson1996}, which involves running a loop-erased random walk on the vertices of $\calG[\calA].$
They differ only in how they partition the spanning tree.
In \texttt{gs2}, we delete a uniformly selected edge from the spanning tree whereas in \texttt{gs3}, we select an edge with probability proportional to the size of the smallest cluster that results when that edge is deleted.
Compared to \texttt{gs2}, \texttt{gs3} is biased towards creating balanced partitions.
Our last strategy (\texttt{gs4}) computes the Fiedler partition of the uniformly drawn spanning tree.}

\subsection{How flexBART ``borrows strength'' across categorical levels}
\label{sec:network_kernel}

\revised{
Recall that the BART prior in Section~\ref{sec:cgm98_prior} induces a prior over partitions of the levels of a categorical predictor.
When the predictor displays network structure, that prior places considerable probability on disconnected network partitions (Figure~\ref{fig:philly_uniform_unordered}).
By modifying the BART prior to use one of \texttt{gs1}, \ldots, \texttt{gs4} when splitting on a network-structured predictor, we force the corresponding prior over network partitions to concentrate only on those partitions with connected components.
Note that the branching process prior over the tree structure $T$ determines the distribution over the number of clusters in the partition and the specific process of drawing $\cutset \subset \calA$ determines the size and composition of the clusters.
More importantly the way we draw $\cutset$ determines how flexBART ``shares information'' across categorical levels.}

\revised{To see this, suppose we (i) have one categorical predictor $X \in \{c_{1}, \ldots, c_{K}\};$ (ii) model $Y \sim \mathcal{N}(f(X), \sigma^{2});$ and (iii) place a BART prior on $f.$
Since $X$ is discrete, estimating $f$ amounts to a normal means problem in which we must estimate $\boldsymbol{\theta} = (\theta_{1}, \ldots, \theta_{K})$ where $f(c_{k}) = \theta_{k}$ for all $k.$
Thanks to the normal prior on the jumps in each tree leaf, it turns out that the induced marginal prior of $\boldsymbol{\theta}$ is multivariate normal and that the prior covariance matrix of $\boldsymbol{\theta}$ determines how estimates of the $\theta_{k}$ are shrunk towards each other.
Using the argument from Section 5.2 of \citet{Linero2017_review}, in the infinite tree limit, the $(k,k')$ entry in the prior covariance matrix of $\boldsymbol{\theta}$ is proportional to the probability that $c_{k}$ and $c_{k'}$ are assigned to the same leaf in a tree. 
So for large ensembles, this co-clustering probability matrix approximately dictates how information is shared across categorical levels.}

\begin{figure}[h!]
\centering
\begin{subfigure}{0.32\textwidth}
\centering
\includegraphics[width = \textwidth]{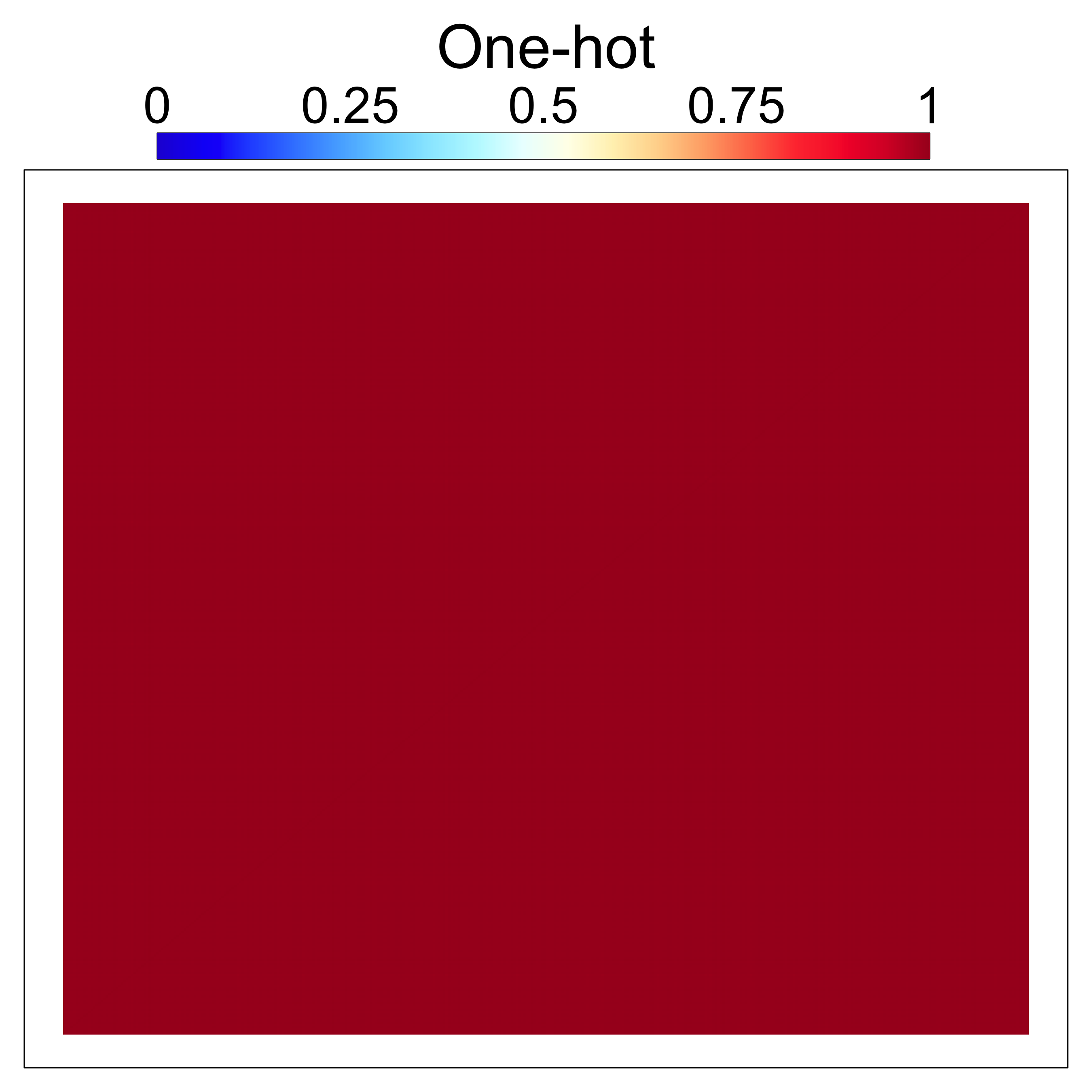}
\caption{}
\label{fig:kernel_remove}
\end{subfigure}
\begin{subfigure}{0.32\textwidth}
\centering
\includegraphics[width = \textwidth]{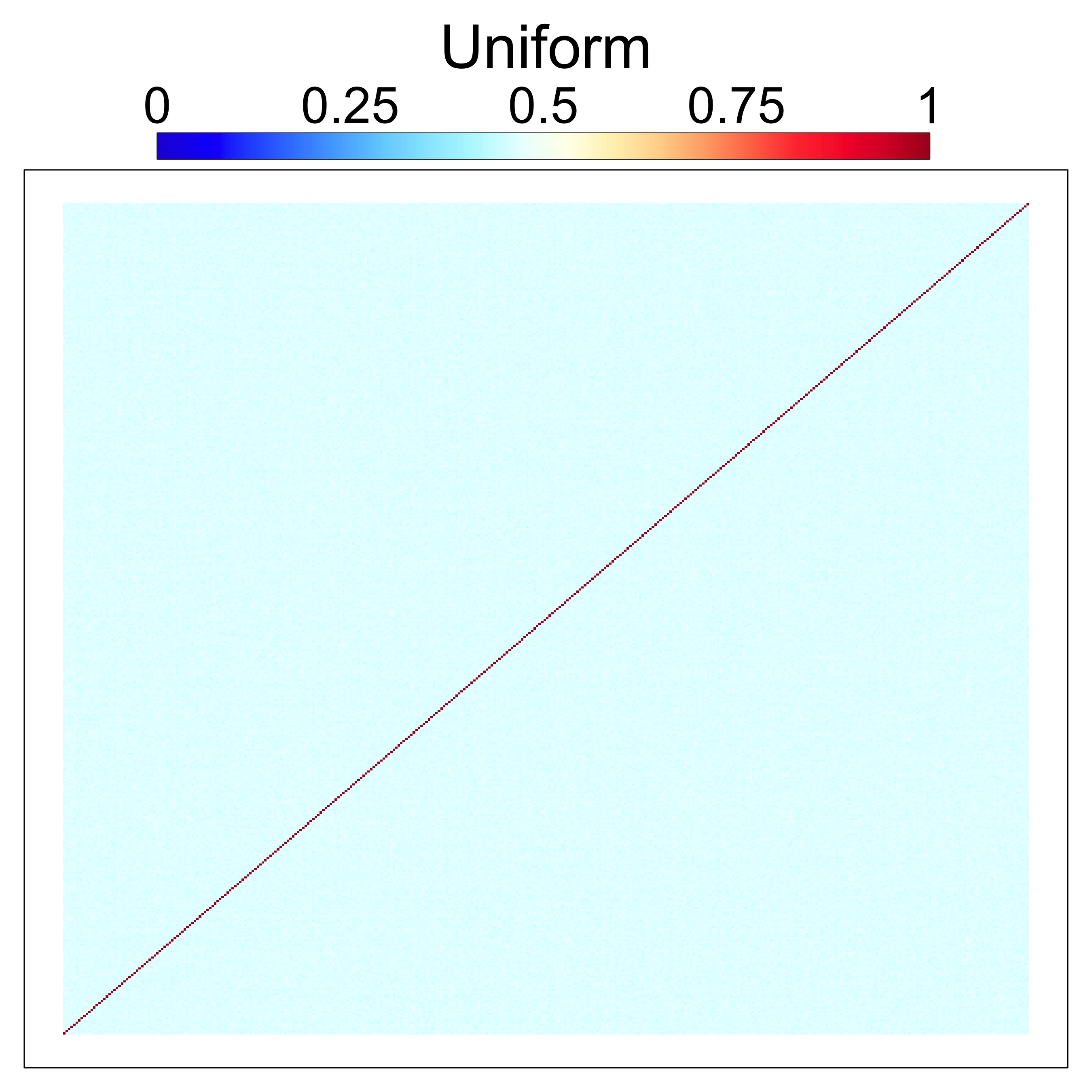}
\caption{}
\label{fig:kernel0}
\end{subfigure}
\begin{subfigure}{0.32\textwidth}
\centering
\includegraphics[width = \textwidth]{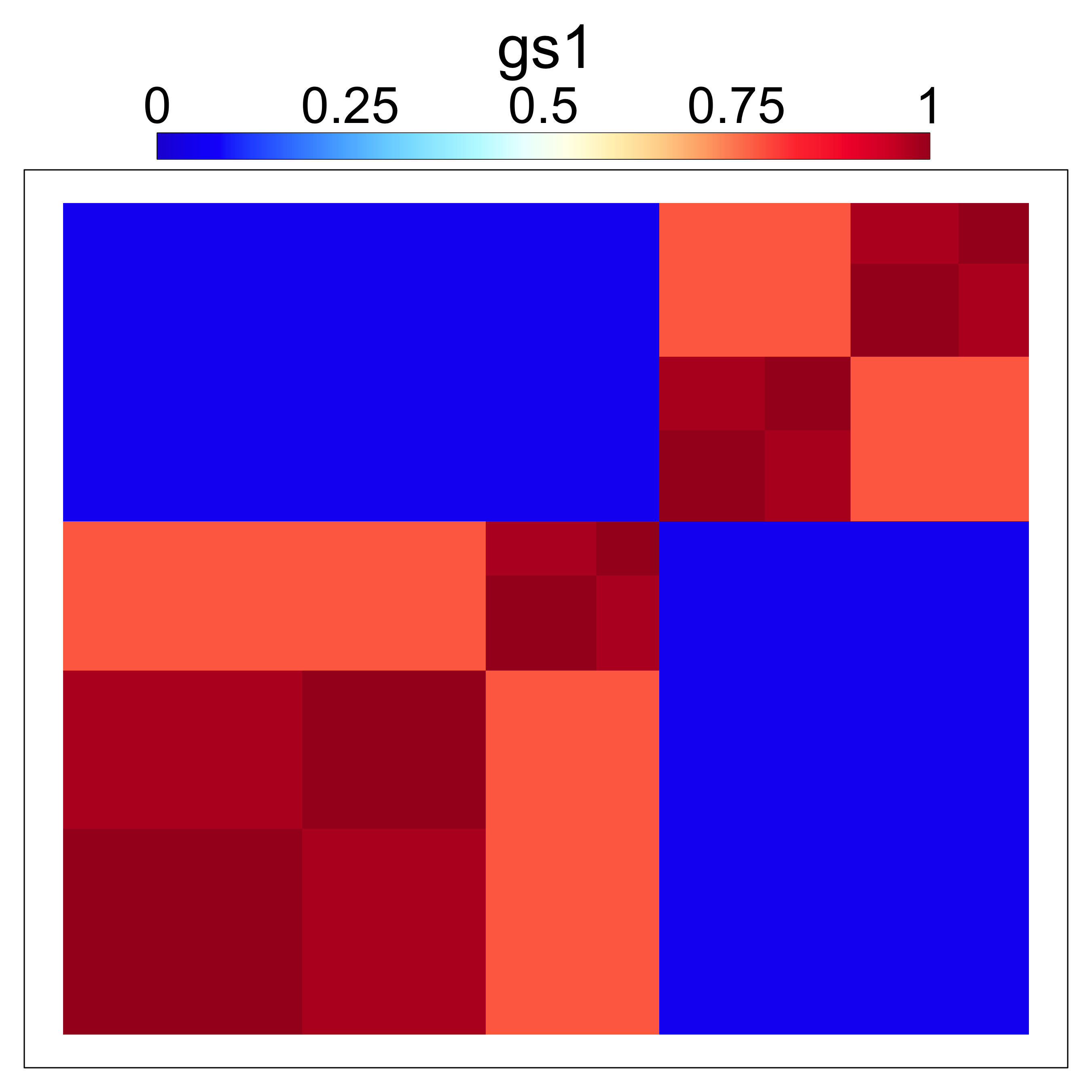}
\caption{}
\label{fig:kernel_gs1}
\end{subfigure}

\begin{subfigure}{0.32\textwidth}
\centering
\includegraphics[width = \textwidth]{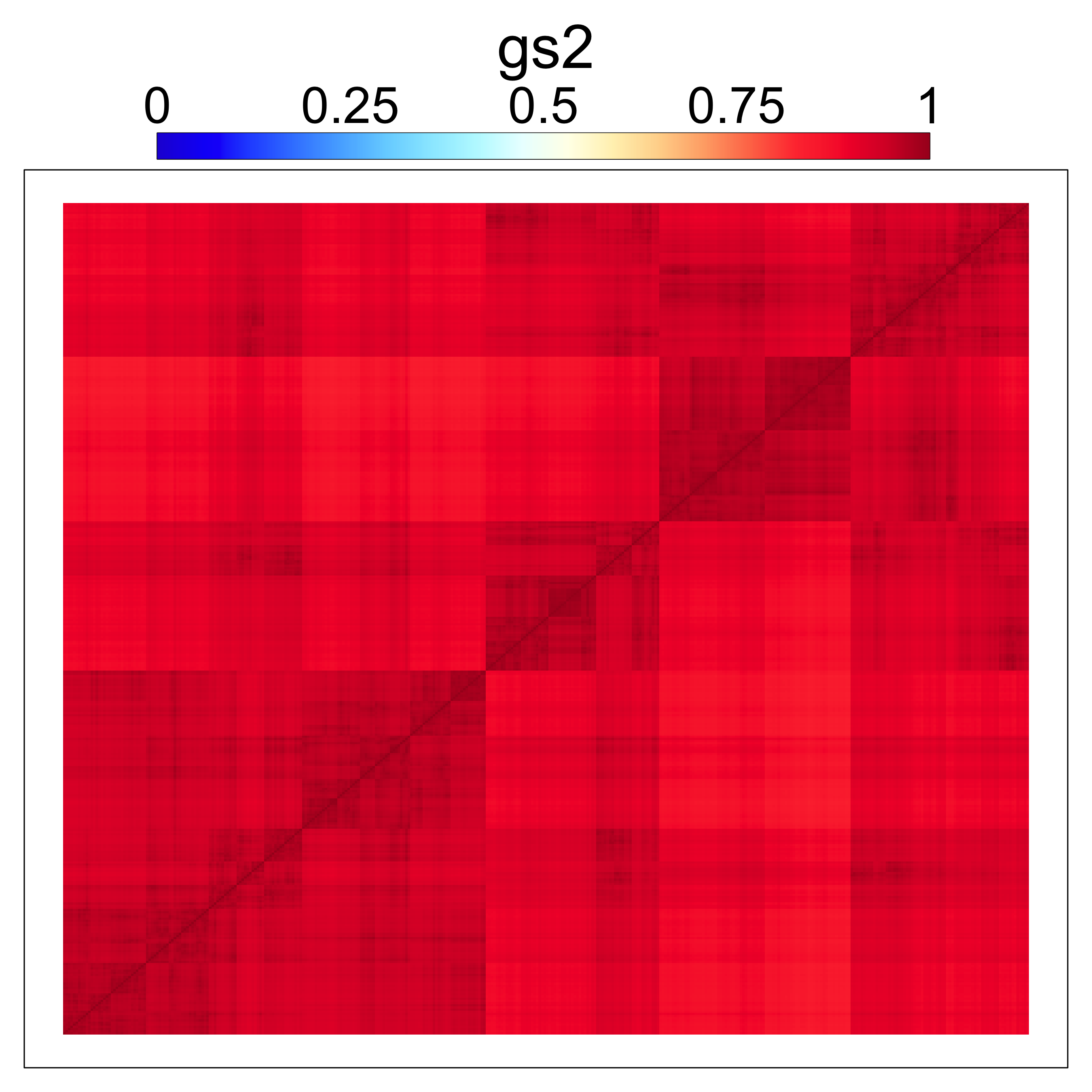}
\caption{}
\label{fig:kernel_gs3}
\end{subfigure}
\begin{subfigure}{0.32\textwidth}
\centering
\includegraphics[width = \textwidth]{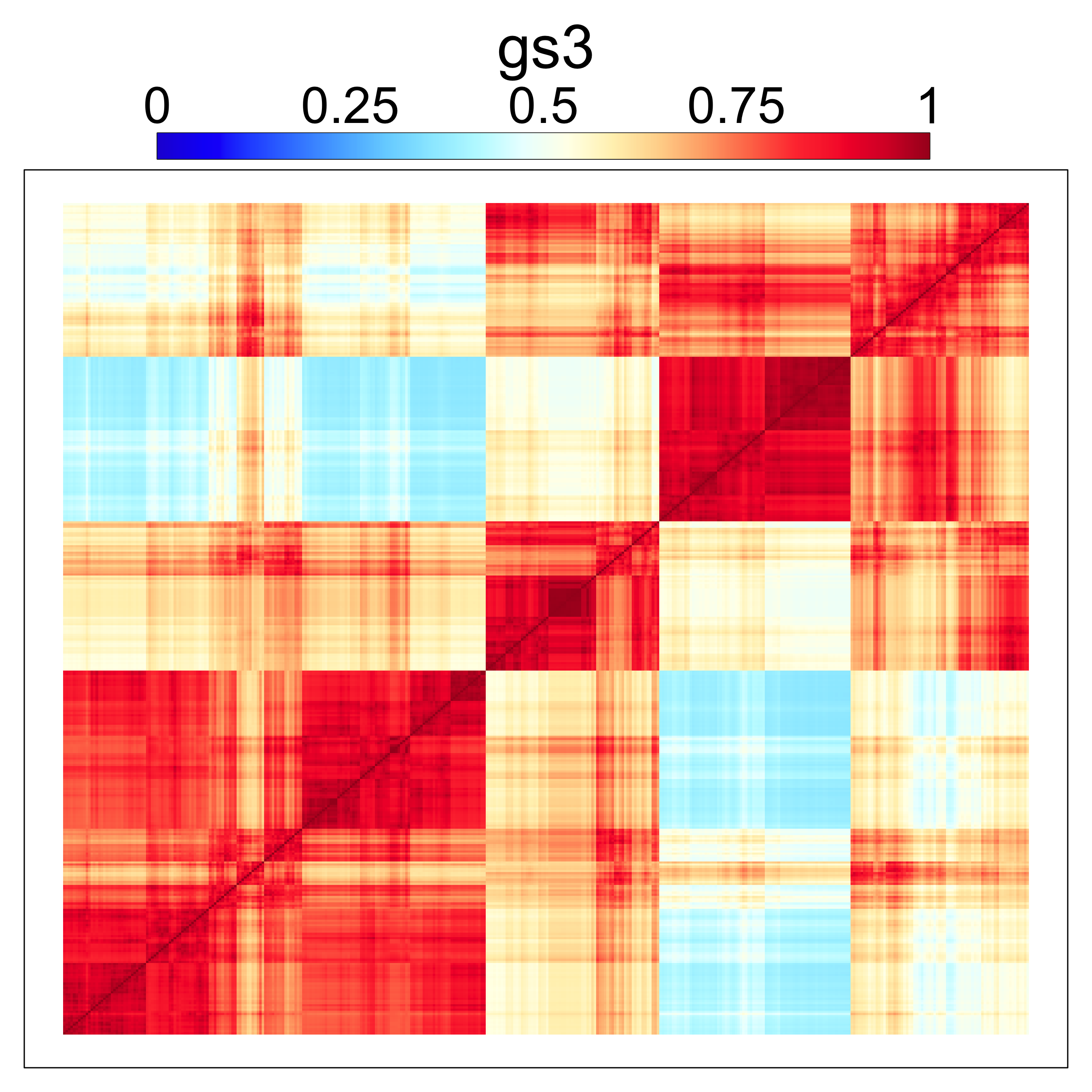}
\caption{}
\label{fig:kernel_gs5}
\end{subfigure}
\begin{subfigure}{0.32\textwidth}
\centering
\includegraphics[width = \textwidth]{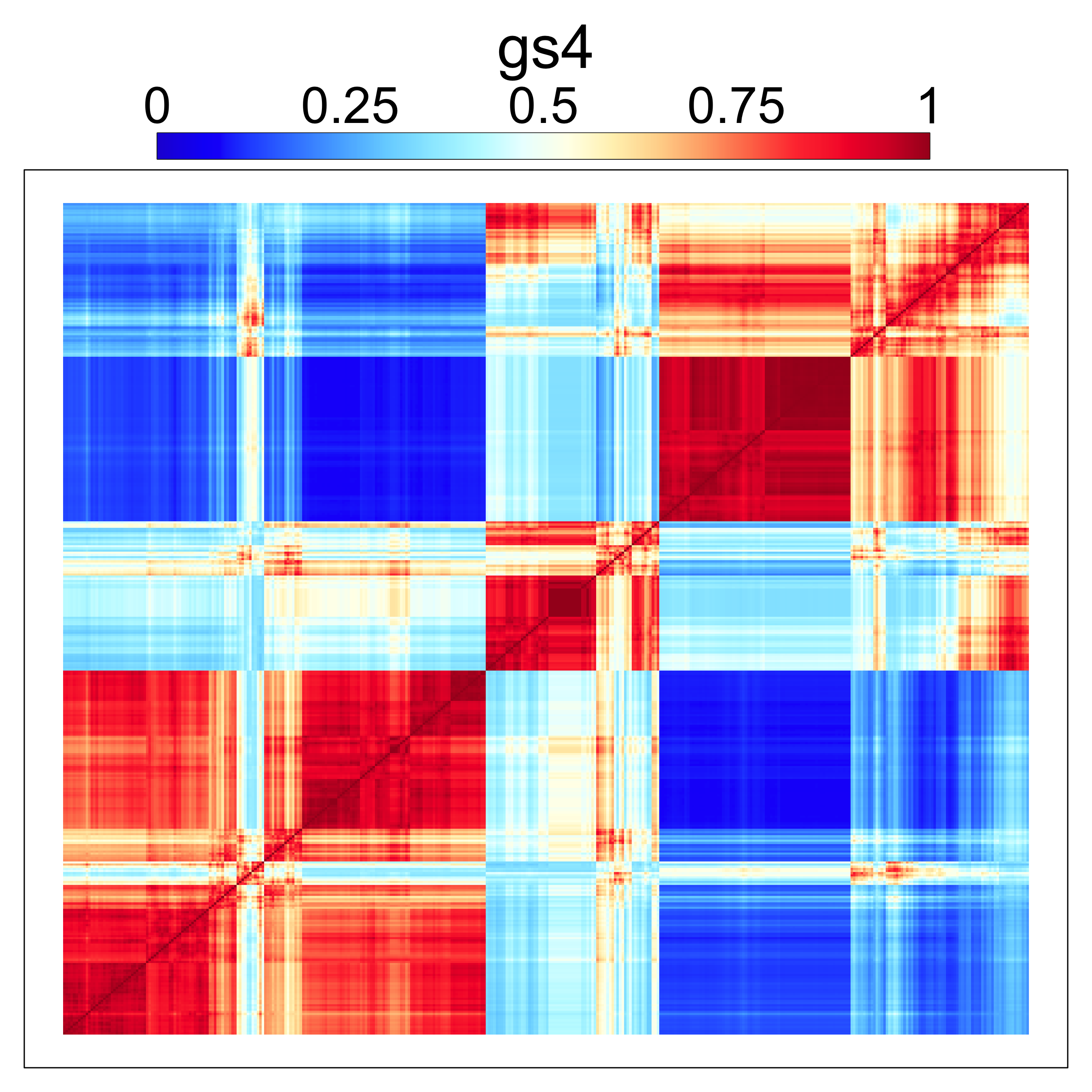}
\caption{}
\label{fig:kernel_gs6}
\end{subfigure}
\caption{Pairwise co-clustering probabilities for the Philadelphia census tracts induced by (i) the standard BART prior with one-hot encoding; (ii) the standard BART prior with naive splits; and (iii) our four proposed network-splitting strategies. Note that the tracts have been arranged so that the co-clustering probabilities from the deterministic \texttt{gs1} displays block structure}
\label{fig:network_kernels}
\end{figure}

\revised{Figure~\ref{fig:network_kernels} visualizes the co-clustering probabilities of the Philadelphia census tracts computed using 10,000 prior draws from the (i) standard BART prior after one-hot encoding; (ii) the standard BART prior that independently splits categorical levels uniformly; and (iii) the BART prior equipped with each of our network-splitting strategies.}

\revised{Recall that one-hot encoding induces partitions by removing one level from the bulk at a time, creating many singleton clusters and one large cluster, which manifests in the extremely large co-clustering probabilities in Figure~\ref{fig:kernel_remove}
In sharp contrast, independently partitioning levels to the left and right branches results in the overall much lower co-clustering probabilities in Figure~\ref{fig:kernel0}.
The symmetry in Figure~\ref{fig:kernel0} reflects the fact that every level is equally likely to be assigned to the left or right.
}

\revised{Because \texttt{gs1} is deterministic, any two trees with the same structure induce exactly the same partition of the census tracts.
As a result, certain blocks of tracts always clustered together while others are never co-clustered.
Such prior rigidity is rather unappealing, since it prevents BART from ever pooling across certain pairs of tracts, irrespective of the data.}

\revised{In sharp contrast, the co-clustering probabilities observed with \texttt{gs2}, which repeatedly draws a uniform spanning tree and then deletes a uniform edge, are all extremely large.
On further inspection, at least for the Philadelphia census tract network, we discovered that the BART prior equipped with \texttt{gs2} tended to produce a large number of singleton clusters.
By varying the way we selected an edge for removal, we observe more variety in the co-clustering probabilities for \texttt{gs3} (Figure~\ref{fig:kernel_gs5}) and \texttt{gs4} (Figure~\ref{fig:kernel_gs6}).}


\revised{To summarize, to fit BART models with network-structured categorical predictors, we modified the original regression tree prior from Section~\ref{sec:cgm98_prior} with more nuanced decision rule prior.
As before, to draw a decision rule $\{X_{j} \in \cutset\}$ at a given tree node, our new prior works by (i) selecting the decision variable index $j$ uniformly at random; (ii) computing the set of values $\calA$ available to $X_{j}$ at the current node; and (iii) setting $\cutset$ equal to a random subset of $\calA.$
And when $X_{j}$ is continuous or an unstructured categorical predictor, our new prior draws $\calC$ exactly as described in Section~~\ref{sec:cgm98_prior}.
However, when $X_{j}$ is a network-structured categorical predictor, our new prior now forms $\calC$ using one of \texttt{gs1}--\texttt{gs4} instead of independently assigning each level in $\calA$ to $\calC.$}

\revised{For posterior computation, we utilize exactly the same Gibbs sampling strategy, in which we sequentially update regression trees conditionally on all others.
We similarly update each regression tree in two steps, first updating the tree structure via Metropolis-Hastings with the grow/prune kernel and then conditionally updating the jumps in each leaf.
But now, whenever we attempt to grow a tree, we draw a decision rule from our new decision rule prior.}

\section{Simulation studies}
\label{sec:simulations}
\revised{We performed several synthetic data experiments comparing our proposed implementation \textbf{flexBART} to that of the default implementation available in the \textsf{R} package \textbf{BART}. 
In Section~\ref{sec:grouping_advantage}, we probe the potential benefits and pitfalls of independently assigning multiple categorical levels to each branch of a tree relative to one-hot encoding.
We then assessed how well BART equipped with each of our network-splitting strategies recovers functions defined over networks in Section~\ref{sec:network_sims}. }

We ran all experiments on a shared high-throughput computing cluster \citep{chtc}.
All results are based on simulating a single Markov chain for 2,000 iterations.
Although we would simulate more chains in actual applied practice, a single chain sufficed for our comparative study. 
Code to reproduce our experiments is available at \url{https://github.com/skdeshpande91/flexBART}.

\subsection{The benefits of pooling multiple levels}
\label{sec:grouping_advantage}

\revised{
We constructed four different data generating processes (DGPs) in which the regression function varied considerably with respect to the value of a categorical predictor.
For each DGP $d \in \{1,2, 3, 4\}$ and $n \in \{1000, 5000, 10000\},$ we generated 50 training datasets $(\bx_{1}, y_{1}), \ldots, (\bx_{n}, y_{n})$ where $y_{i} \sim \normaldist{\mu_{d}(\bx_{i})}{1}$ and each $\bx_{i}$ was drawn uniformly from $[0,1]^{10} \times \{c_{1}, \ldots, c_{10}\}.$}
\revised{We then assessed how well different implementations of BART could estimate the regression function $\mu_{d}(\bx)$ at 500 new $\bx$ values.
We built the four regression functions $\mu_{1}(\bx), \ldots, \mu_{4}(\bx)$ using a common set of basis functions:
\begin{align*}
f_{0}(\bx) &= 10 \sin\left(\pi x_{1}x_{2}\right), \quad f_{1}(\bx) = 10(x_{3} - 0.5)^{2},  \quad f_{2}(\bx) = 10(x_{3} - 0.5)^{2} + 10x_{4} + 5x_{5}, \\
f_{3}(\bx) &= 6x_{1} + (4 - 10 \times \ind{x_{2} > 0.5})\times\sin(\pi x_{1}) - 4\times \ind{x_{2} > 0.5} + 15 & ~ & ~ 
\end{align*}
We specifically set
\begin{align*}
\mu_{1}(\bx) &= (f_{0}(\bx) + f_{1}(\bx) + f_{2}(\bx) - 0.75) \times \ind{x_{11} \not\in \{c_{0}, c_{2}, c_{4}, c_{8}\}}+ f_{3}(\bx) \times \ind{x_{11} \in \{c_{0}, c_{2}, c_{4}, c_{8}\}} \\
\mu_{2}(\bx) &= (f_{0}(\bx) + f_{1}(\bx) + f_{2}(\bx) - 0.75) \times \ind{x_{11} = c_{0}} + f_{2}(\bx) \times \ind{x_{11} \neq c_{0}} \\
\mu_{3}(\bx) &= f_{0}(\bx) \times \ind{x_{11} \in \{c_{0}, c_{3}, c_{4}, c_{6}\}} + f_{1}(\bx) \times \ind{x_{11} \in \{c_{1}, c_{3}, c_{4}, c_{5}, c_{6}\}} \\
~&~~+ f_{2}(\bx) \times \ind{x_{11} \in \{c_{2}, c_{3}, c_{5}, c_{6}\}} + f_{3}(\bx) \times \ind{x_{11} \in \{c_{7}, c_{8}, c_{9}\}} \\
\mu_{4}(\bx) &= \sum_{\ell = 0}^{9}{\ind{x_{11} \in c_{\ell}} \times \left[\frac{\ell + 1}{10} \times (f_{1}(\bx) + f_{2}(\bx) + f_{3}(\bx)-0.75) + \frac{9-\ell}{10} \times f_{3}(\bx)\right]}
\end{align*}
}

\revised{In DGP1 and DGP2, the regression function is identical across two groups of categorical levels.
However, in DGP1, both groups contain multiple levels whereas in the DGP2, one group contains a single level. 
In DGP3 and DGP4, although the relationship between $\bx$ and $y$ is different for each level, there is considerable similarity between multiple levels.}

\revised{
We compared \texttt{flexBART}, which assigns categorical levels uniformly to the left and right, to \texttt{BART}, which uses one-hot encoding, and \texttt{targetBART}, which uses target encoding.
Target encoding works by replacing each categorical predictor with a numerical predictor whose entries correspond to the average outcomes in each level.
Many popular implementations of random forests, including those in \texttt{scikit-learn} \citep{scikit-learn} and the \textsf{R} package \textbf{ranger} \citep{ranger}, rely on target encoding.
Constructing new predictors from the response, however, is arguably pathological.}

\revised{We additionally compared \texttt{flexBART}, \texttt{BART}, and \texttt{targetBART} to an oracle procedure (\texttt{oracleBART}) that divides the training data according to the true partition underlying $\mu_{d}(\bx)$ and fits separate BART models to each part.
For DGP 1 the oracle fits two separate BART models, one for observations with $x_{11} \in \{c_{0}, c_{2}, c_{4}, c_{8}\}$ and one for observations with $x_{11} \not\in \{c_{0}, c_{2}, c_{4}, c_{8}\}.$
The oracle similarly fits two separate BART models for DGP2, one for observations with $x_{11} = c_{0}$ and one for observations with $x_{11} \neq c_{0}.$
For DGP3, the oracle fits eight BART models, one each for levels $c_{0}, \ldots, c_{6}$ and one for levels $c_{7}, c_{8},$ and $c_{9}.$
Finally, the oracle fits a separate BART model to each level in DGP4. }

\revised{For brevity, we only report the results for $n = 5000.$
The results for other values of $n$ are qualitatively similar and are reported in Appendix~\ref{app:synthetic}.
Figure~\ref{fig:grouping_advantage_n5000} compares the out-of-sample mean square errors of all methods relative to \texttt{BART}.
Values less than one in the figure represent better predictive accuracy than \texttt{BART}.}

\begin{figure}[ht]
\centering
\includegraphics[width = 0.8\textwidth]{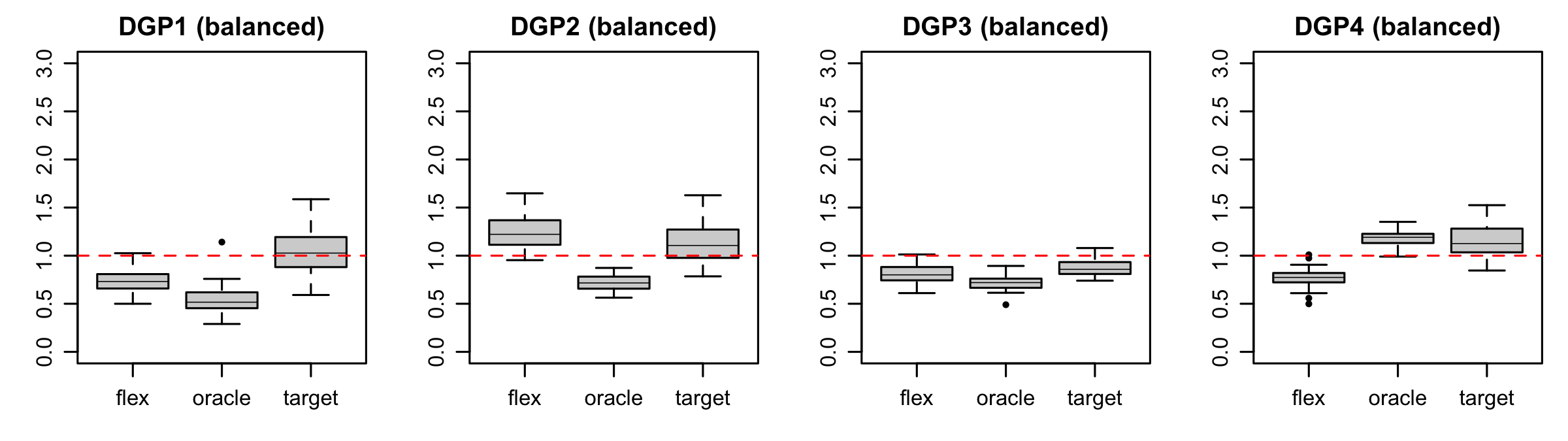}
\caption{Out-of-sample mean square errors relative to \texttt{BART} for all data generating process with $n = 5000$ training observations. Values less (resp.\ greater) than one indicate better (resp.\ worse) performance than \texttt{BART}.}
\label{fig:grouping_advantage_n5000}
\end{figure}

\revised{Except for DGP2, \texttt{flexBART} consistently outperformed \texttt{BART}: the average \texttt{flexBART} out-of-sample error was 18\% (DGP1), 13\% (DGP3), and 20\% (DGP4) lower than \texttt{BART}'s and \texttt{flexBART}'s errors were smaller than \texttt{BART}'s across nearly all simulation replications.
Across these three DGPs, the differences in predictive performance between \texttt{flexBART} and \texttt{BART} were statistically significant, even after accounting for multiple comparisons; see Table~\ref{tab:grouping_advantage}.}

\revised{The results for DGP1 highlight the benefits of pooling together multiple categorical levels: \texttt{BART} was unable to effectively separate levels $\{c_{0}, c_{2}, c_{4}, c_{8}\}$ from the others and often attempted to pool data across these two different groups.
Interestingly, \texttt{targetBART} often performed \emph{worse} than \texttt{BART} for this DGP.
Recall that \texttt{targetBART} orders the levels $c_{\ell}$ according to the average responses and partitions the levels in a manner consistent with that ordering.
For DGP1, this ordering was often inconsistent with the true underlying partition, leading \texttt{targetBART} to inappropriately pool data across levels with different regression functions.}

\revised{Although the regression function varied across levels $c_{0}, \ldots, c_{6}$ in DGP3, the partial sum structure of $\mu_{3}(\bx)$ induced correlation across these levels.
\texttt{flexBART} adapted to this dependence structure more effectively than \texttt{BART}.
Similarly, \texttt{flexBART} leveraged the similarities across levels in DGP4 better than \texttt{BART}.
Interestingly, \texttt{flexBART} outperformed the oracle, which fit separate BART models to the data in each level, in DGP4.}

\revised{The partition of $X_{11}$'s levels underpinning DGP2 consists of one singleton set and one set with the other nine levels.
Because it assigns categorical levels to the left or right with equal probability, \texttt{flexBART} is biased towards balanced partitions and is generally unable to isolate a single outlying level.
\texttt{BART}, in contrast, always separates one level from the bulk at each decision node.
On the view that the prior over partitions of $X_{11}$'s levels induced by \texttt{BART} is better aligned with the actual partition underlying DGP2 than the prior induced by \texttt{flexBART}, \texttt{flexBART}'s somewhat worse performance relative to \texttt{BART} on DGP2 is not wholly surprising.
Our results suggest that \texttt{flexBART}'s partitioning strategy is preferable to one-hot or target encoding except when one level has a substantially different, outlying response surface than the others.
The same conclusions hold even when there is severe imbalance in the relative frequencies of $X_{11}$'s levels; see Figure~\ref{fig:grouping_advantage_imbal}.}

\revised{Across all simulation replications, \texttt{flexBART} was substantially faster than \texttt{BART} and \texttt{targetBART}.
Averaging across all DGPs, \texttt{flexBART} took 14 seconds, ($n = 1000$), 27 seconds ($n = 5000$), and 47 seconds ($n = 10000$) to simulate 2,000 Markov chain iterations.
To draw the same number of samples, \texttt{BART} and \texttt{targetBART} took 20 seconds ($n = 1000$), 75 seconds ($n = 5000$), and 150 seconds ($n = 10000$).
\texttt{flexBART}'s speedup is a byproduct of our implementation, which removes certain redundant calculations performed in the \textbf{BART} codebase; see Appendix~\ref{app:implementation_details}.}

\subsection{Network-linked regression}
\label{sec:network_sims}

\revised{We performed two simulation studies to determine flexBART's sensitivity to the choice of network-splitting strategy.
In these experiments, we additionally compared flexBART to a hybrid method that first uses the adjacency spectral embedding \citep[ASE;][]{Sussman2012} to embed the network vertices into a latent space and then uses the embeddings as continuous predictors in a BART model.
Doing so allowed us to compare direct (via network partitioning) and indirect (via network embedding) incorporation of network structure in BART models.} 

\revised{To briefly describe the ASE, let $A$ be a network adjacency matrix and suppose that its singular value decomposition is given by $A = U\Sigma V^{\top},$ where $\Sigma$ has decreasing main diagonal.
The $d$-dimensional ASE is given by $\hat{U}_{d}\hat{\Sigma}_{d}^{1/2}$ where $\hat{U}_{d}$ contains just the first $d$ columns of $U$ and $\hat{\Sigma}_{d}$ the truncation of $\Sigma$ to its first $d$ rows and columns.
Recently, \citet{Lunde2023} and \citet{Hayes2023} have incorporated low-dimensional ASEs as features in regression models.
In our experiments, we compared flexBART equipped with each of \texttt{gs1}--\texttt{gs4} to (i) BART that includes the $d$-dimensional ASE for $d \in \{1,3,5\}$ as additional predictors; (ii) BART that one-hot encodes network labels (\texttt{BART}); and (iii) flexBART that uniformly assigns levels to the left and right branch (\texttt{flexBART\_unif}).}

\revised{In both experiments, we generated $t = 100$ noisy observations of a function defined over a subset of vertices in the Philadelphia census tract network.
For the first experiment, we used a piecewise constant function and for the second experiment, we used a function that smoothly interpolated between two functions.
Specifically, at vertex $v,$ we generated noisy evaluations of $g(\bx, v) = w_{v}g_{0}(\bx) + (1 - w_{v})g_{1}(\bx),$ where the vertex weight $w_{v}$ varied smoothly from 0 to 1 (see Figure~\ref{fig:network_dgp}) and 
\begin{align*}
g_{0}(\bx) &= 3x_{1} + (2 - 5 \times \ind{x_{2} > 0.5})\times\sin(\pi x_{1}) - 2\times \ind{x_{2} > 0.5}\\
g_{1}(\bx) &= 3 - 3\times \cos(6\pi x_{1}) \times x_{1}^{2} \times \ind{x_{1} > 0.6} - 10 \times \sqrt{x_{1}} \times \ind{x_{1} < 0.25}.
\end{align*}}
\revised{
In both experiments, we trained our BART models using data from only 90\% of all vertices and compared each method's ability to evaluate the function at the held-out vertices.
Note that we did not delete the held-out vertices and their incident edges from the network while training.
In a sense, our experiments probe the ability of each BART implementation to interpolate at an existing network vertex rather than extrapolate to entirely new vertices. }

\revised{Unsurprisingly, methods that did not account for network structure performed substantially worse than methods that did.
The out-of-sample root mean square error (RMSE) for \texttt{BART} and \texttt{flexBART\_unif} were 39.036 and 29.791 for the piecewise constant function and 0.953 and 1.866 for the smoothly varying function.
By comparison, the RMSEs for flexBART equipped with our network-splitting strategies were 6.995 (\texttt{gs1}), 3.708 (\texttt{gs2}), 4.471 (\texttt{gs3}), and 4.384 (\texttt{gs4}) for the piecewise constant function.
For the smoothly varying function the respective RMSEs were 0.226, 0.204, 0.191, and 0.204.
Interestingly, despite ostensibly leveraging adjacency information, \texttt{BART\_ase1} performed \emph{worse} than \texttt{BART} on both experiments with RMSEs of 46.459 and 0.977.
Although using higher-dimensional ASEs generally improved predictive performance, both \texttt{BART\_ase3} and \texttt{BART\_ase5} performed worse than any of network-splitting flexBART implementations for the smoothly varying function.
These results, which are available in Table~\ref{tab:network_results}, indicate that directly leveraging adjacency information via stochastic graph partitions yielded significantly better predictions than incorporating adjacency information via embeddings.
Although our network-splitting flexBART implementations repeatedly computed eigendecompositions and ran random walks on the networks, they were still faster than \texttt{BART}, averaging between 152 (\texttt{gs2}) and 220 (\texttt{gs3}) seconds compared to \texttt{BART}'s 258 seconds to draw 2,000 samples.}

\section{Real data results}
\label{sec:real_data}
\subsection{Benchmark datasets}
\label{sec:bakeoff}

\revised{
In Section~\ref{sec:grouping_advantage}, we saw that uniformly partitioning categorical levels to the left and right branches of a tree could lead to substantial improvements over one-hot and target encoding on synthetic data.
We compared the non-oracle methods on sixteen benchmark datasets.
We obtained most of these datasets from the UCI repository (\url{https://archive.ics.uci.edu}); the legacy website for the \emph{Journal of Applied Econometrics} data archive (\url{http://qed.econ.queensu.ca/jae/}); and from several \textsf{R} packages. 
Table~\ref{tab:benchmark} lists the datasets together with their source and dimension.
For each dataset, we formed 50 75\%-25\% training-testing splits.
In the table, we also report the average standardized mean square error (SMSE)\footnote{The SMSE of a prediction method $\hat{Y}$ is defined as the ratio of the average value of $(Y_{\text{test}} - \hat{Y})^{2}$ to the average value of $(Y_{\text{test}} - \overline{Y}_{\text{train}})^{2}.$ $1 - \text{SMSE}$ can be interpreted similarly to $R^{2},$ as a proportion of variation explained} of each method and the average runtime, in seconds.
For each dataset, we performed a one-sided paired t-test comparing the out-of-sample SMSE of \texttt{flexBART} to the other methods.}

\begin{table}[ht]
\centering
\caption{Benchmark datasets, their source and dimensions (category sizes listed in parentheses), and the average standardized mean square error and runtimes (in seconds) for each method. Starred values indicate that \texttt{flexBART} had statistically significantly better performance at the 5\% level, after Bonferonni correction for the 32 comparisons.}
\label{tab:benchmark}
{
\scriptsize
\begin{tabular}{p{4cm}lp{2.2cm}p{1cm}p{1cm}p{1.2cm}p{1cm}p{1cm}p{1.2cm}}
Dataset & $n$ & $\pcont,\pcat$ & \texttt{flexBART} SMSE & \texttt{BART} SMSE & \texttt{targetBART} SMSE & \texttt{flexBART} Time & \texttt{BART} Time & \texttt{targetBART} Time \\ \hline
\texttt{abalone} (\href{https://archive.ics.uci.edu/dataset/1/abalone}{UCI}) & 4177 & 7,1(3) & 0.444 & 0.444 & 0.443 & 15.1 & 43.7 & 40.2 \\
\texttt{ais} (\href{https://rdrr.io/cran/locfit/man/ais.html}{\textbf{locfit}}) & 202 & 10,2(10,2) & 0.118 & 0.115 & 0.119 & 3.5 & 3.9 & 4.1 \\
\texttt{Alcohol} (\href{http://qed.econ.queensu.ca/jae/2001-v16.2/kenkel-terza/}{JAE}) & 2467 & 12,6(6,3,4,3,4,3) & 0.961 & 0.966$^{*}$ & 0.961 & 11.0 & 30.3 & 29.1 \\
\texttt{amenity} (\href{http://qed.econ.queensu.ca/jae/2002-v17.6/chattopadhyay/readme.ch.txt}{JAE}) & 3044 & 23,2(3,4) & 0.288 & 0.289 & 0.288 & 14.2 & 39.4 & 38.1 \\
\texttt{attend} (\href{https://www.rdocumentation.org/packages/UsingR/versions/2.0-7/topics/MLBattend}{\textbf{UsingR}}) & 838 & 6,3(31,2,3) & 0.212 & 0.262$^{*}$ & 0.261$^{*}$ & 7.7 & 9.3 & 10.3 \\
\texttt{cane} (\href{http://www.statsci.org/data/oz/cane.html}{OzDASL}) & 3775 & 22,6(15,5,5,33,29,6) & 0.614 & 0.691$^{*}$ & 0.695$^{*}$ & 14.7 & 31.9 & 29.6 \\
\texttt{Caschool} (\href{https://rdrr.io/cran/Ecdat/man/Caschool.html}{\textbf{Ecdat}}) & 420 & 12,1(45) & 0.002 & 0.005$^{*}$ & 0.005$^{*}$ & 4.6 & 5.8 & 6.0 \\
\texttt{cpu} (\href{https://archive.ics.uci.edu/dataset/29/computer+hardware}{UCI}) & 209 & 6,1(30) & 0.154 & 0.222$^{*}$ & 0.175$^{*}$ & 4.0 & 3.9 & 4.0 \\
\texttt{Engel} (\href{http://qed.econ.queensu.ca/jae/1998-v13.2/delgado-mora/}{JAE}) & 23971& 3,2(5,2) & 0.573 & 0.576$^{*}$ & 0.575$^{*}$ & 88.9 & 253.4 & 256.2 \\
\texttt{fuelEcon} (\url{fueleconomy.gov}) & 20662 & 5,5(6,5,84,33,18) & 0.000 & 0.000 & 0.000 & 75.5 & 338.7 & 266.4 \\
\texttt{Insur} (\href{https://rdrr.io/cran/GLMsData/man/motorins.html}{\textbf{GLMsData}}) & 2182&4,2(7,9) & 0.021 & 0.297$^{*}$ & 0.301$^{*}$ & 16.7 & 23.4 & 23.2 \\
\texttt{Medicare (\href{https://rdrr.io/cran/Ecdat/man/OFP.html}{\textbf{Ecdat}})} &4406&6,8(4,3,2,2,2,2,2,2) & 0.883 & 0.882 & 0.886 & 20.2 & 52.9 & 54.8 \\
\texttt{mpg} (\href{https://archive.ics.uci.edu/dataset/9/auto+mpg}{UCI}) &392&6,1(3) & 0.488 & 0.495$^{*}$ & 0.497$^{*}$ & 3.9 & 5.9 & 5.8 \\
\texttt{servo} (\href{https://archive.ics.uci.edu/dataset/87/servo}{UCI}) &167&2,2(5,5) & 0.153 & 0.189$^{*}$ & 0.158$^{*}$ & 3.6 & 3.8 & 4.0 \\
\texttt{spouse} (\href{http://qed.econ.queensu.ca/jae/1998-v13.5/olson/}{JAE}) & 22272 &11,3(6,3,4) & 0.171 & 0.171 & 0.171 & 85.7 & 269.2 & 285.0 \\
\texttt{strike} (\href{https://lib.stat.cmu.edu/datasets/strikes}{Statlib}) &625& 5,1(18) & 0.307 & 0.329$^{*}$ & 0.313$^{*}$  & 4.0 & 5.2 & 5.6 \\ \hline
\end{tabular}
}
\end{table}
\revised{
For all but four datasets (\texttt{ais}, \texttt{Medicare}, \texttt{fuelEcon}, and \texttt{spouse}), \texttt{flexBART} had smaller average out-of-sample mean square error than \texttt{BART}. 
Looking more closely, the gaps between \texttt{flexBART} and \texttt{BART} on these datasets were extremely small.
For instance, \texttt{BART}'s SMSE of 0.115 on \texttt{ais} suggests that it could explain about 88.5\% of the variation in the outcome, which was only slightly better than \texttt{flexBART}, which could only explain about 88.2\% of the variation.
The differences in SMSE (equiv.\ variation explained) for \texttt{Medicare} and \texttt{spouse} were similarly small (0.001 and 0.0005).
And for \texttt{fuelEcon}, both \texttt{BART} and \texttt{flexBART}  predicted the response almost perfectly (SMSEs of 0.0002 and 0.0003), rendering the performance gap essentially negligible.}

\revised{Of the 12 datasets in which \texttt{flexBART} outperformed \texttt{BART}, the improvement was statistically significant in 10 datasets.
While some of the improvements were quite modest (e.g., \texttt{flexBART} was about 0.5\% better for \texttt{Engel}), others were quite substantial (e.g., a 19\% improvement on \texttt{servo} and a 30\% improvement on \texttt{cpu}).
It is notable that the largest improvements occurred on datasets where axis-aligned \texttt{BART} already produced reasonably good predictions.
For instance, while \texttt{BART} respectively accounted for about 78\%, 70\%, and 80\% of the variability on \texttt{cpu}, \texttt{Insur}, and \texttt{servo}, \texttt{flexBART} respectively accounted for 84\%, 98\%, and 85\%. 
Conversely, however, when \texttt{BART} provided rather poor fits to the data (\texttt{Alcohol} and \texttt{Medicare}), \texttt{flexBART} did not offer dramatic improvements. 
In other words, while \texttt{flexBART} can make a good model better (and sometimes substantially so), it cannot save a bad model. 
Interestingly, the scale of \texttt{flexBART}'s improvements did not closely track with the number of categorical predictors nor with the number of categorical levels.
\texttt{flexBART} similarly compares favorably to \texttt{targetBART}. 
\texttt{flexBART} tended to be much faster than both \texttt{BART} and \texttt{targetBART}, with the gap widening for larger values of $n.$}

\subsection{Pitch framing}
\label{sec:pitch_framing}

We scraped pitch-by-pitch data from each Major League Baseball season between 2013 and 2019 using the \textbf{baseballr} package \citep{baseballr_package}.
Letting $y$ be a binary indicator of ball ($y = 0$) or strike ($y = 1$) and concatenating pitch location and the identities of the players and umpire into the vector $\bx,$ we modeled $\P(y = 1) = \Phi(f(\bx))$ where $\Phi$ is the standard normal cumulative distribution function and used BART to approximate $f.$
For each season, we generated 10 90\%-10\% training-testing splits and computed the posterior mean called strike probability for each pitch in every training and testing dataset. 
We assessed the predictive performances of \revised{\texttt{flexBART}, \texttt{BART}, and \texttt{targetBART}} using misclassification rate, \revised{log-loss, and Brier score (i.e., mean square error). For brevity, we only report results on misclassification rates here and defer the qualitatively similar results for the other metrics to Appendix~\ref{app:pitch_framing}.}

\revised{
Both \texttt{flexBART} (8.6\%) and \texttt{targetBART} (8.8\%) achieved smaller out-of-sample misclassification rates than the 10.6\% reported in Table 2 of \citet{DeshpandeWyner2017}.
Surprisingly, the one-hot encoded \texttt{BART} performed substantially \emph{worse}, with an average misclassification rate of 18.4\% across the seven seasons worth of data.
Across every cross-validation fold in every season \texttt{flexBART} had significantly smaller misclassification error than \texttt{BART} (one-sided p-value = $4.5\times10^{-26}$) and \texttt{targetBART} (one-sided p-value = $2.7\times10^{-22}$).
Our results suggest that \emph{how} one accounts for the categorical player and umpire identifies makes a material difference in predictive quality: whereas one-hot encoding these identifies yields substantially worse predictions than \citet{DeshpandeWyner2017}'s model with no umpire-player or player-location interactions, the more flexible partitioning performed by \texttt{flexBART} produced much better predictions.
\texttt{flexBART} took an average of 48 minute to draw 2,000 posterior samples, which was faster than the two hours required by \texttt{BART}.}

\subsection{Philadelphia crime data}
\label{sec:philly_crime}

We obtained our crime data from \url{opendataphilly.org}.
For each census tract, following  \citet{Balocchi2022_crime}, we computed the monthly crime density, defined as number of crimes per square mile, and applied an inverse hyperbolic sine transformation to counteract the considerable skewness. 
Letting $y_{v,t}$ be the transformed crime density in census tract $v$ at time $t,$ with $t = 1$ corresponding to January 2006 and $t = 192$ corresponding to December 2021, we modeled $y_{v,t} \sim \normaldist{f(t,v)}{\sigma^{2}}$ and used BART to estimate $f(t,v).$

\revised{We created 100 75\%-25\% training-testing splits and compared the methods from Section~\ref{sec:network_sims} in terms of out-of-sample prediction.
With the exception of \texttt{BART\_ase1}, which had an average RMSE of 4.05, most methods provided reasonably good out-of-sample predictions, with RMSEs ranging from 1.21 (\texttt{gs2}) to 1.93 (\texttt{gs1}).
For context, the standard deviation of the $y_{vt}$'s was about 4.65, meaning that all methods except \texttt{BART\_ase1} were able to estimate at least 84\% of the variability in the testing set.
Like in our synthetic experiments, accounting for network structure via the deterministic \texttt{gs1} or the one-dimensional ASE (\texttt{BART\_ase1}) yielded substantially worse performance than ignoring network structure entirely: the average RMSEs were about 1.32 (\texttt{BART}), 1.33 (\texttt{targetBART}), 1.23 (\texttt{flexBART\_unif}), 1.93 (\texttt{gs1}), and 4.05 (\texttt{BART\_ase1}).}

\revised{Our stochastic network-splitting strategies all yielded substantially better out-of-sample RMSEs than both \texttt{BART\_ase3} (1.90) and \texttt{BART\_ase5} (1.57): the RMSEs were 1.21 (\texttt{gs2}, \texttt{gs3}), and 1.25 (\texttt{gs4}).
These results again highlight the merits of direct incorporation of network structure via partitioning over indirect incorporation via embedding.
Although the gap between \texttt{gs2} and \texttt{flexBART\_unif}, which does not account for network structure, is small (RMSEs of 1.21 and 1.23), this difference is statistically significant (two-sided p-value = $2.8 \times 10^{-52}$).
Moreover, \texttt{gs2}'s error was smaller than \texttt{flexBART\_unif}'s in every simulation replication.}

\section{Discussion}
\label{sec:discussion}
Existing implementations of BART one-hot encode categorical predictors, operationalizing them as several binary indicators.
By adopting such a representation, they effectively specify a prior that places zero probability on the vast majority of partitions of the levels, limiting their ability to partially pool data across categories.
We overcome this limitation with a new implementation of BART using a more general class of regression trees. 
In our trees, multiple levels of a categorical variable can be assigned to both the left and right child of decision nodes.
BART ensembles built using our new trees can borrow strength much more flexibly across categorical levels.
\revised{Using several synthetic datasets and real-world benchmarks, we found that our new implementation often yielded more accurate predictions than implementations using one-hot encoding.
Specifically, our results suggest that (i) flexBART can improve an already good one-hot encoded BART model; (ii) that this improvement can be substantial; but (iii) flexBART cannot save an extremely poorly performing one-hot encoded BART model.
We further found that flexBART often outperforms a version of target-encoded version of BART.}

Motivated by a spatial analysis of crime in Philadelphia, we introduced \revised{several} stochastic processes that, when incorporated into our decision rule prior, enables our regression trees to form spatially contiguous partitions of small geographic regions.
\revised{Equipping BART with these processes, which directly leverage network structure via partitioning, often produced much better predictions than indirectly incorporating network structure via network embeddings.
We recommend using either \texttt{gs2} or \texttt{gs3} as a default.}

While we focused on \revised{spatial} data aggregated within census tracts, we can seamlessly model variation at multiple spatial resolutions using flexBART.
\revised{For instance, one could pass multiple network-structured categorical predictors encoding membership in a series of nested geographic areas like census blocks, block groups, and tracts.}
\revised{Beyond spatial categorical predictors, \textbf{flexBART} can be extended to handle categorical predictors with more complicated structure like a partial order or a hierarchical organization.
For partially ordered levels, one could use the network-splitting proposals introduced here to partition the Hasse diagram encoding the partial order relationships between the levels.
For hierarchically organized levels, one could arrange the levels into a dendrogram and embed a procedure for recursively cutting the dendrogram's branches into the decision rule prior.
We leave such extensions to future work.}

More substantively, our argument against one-hot encoding categorical predictors when fitting BART models is based on a particular lack-of-support phenomenon.
It is important to note that the cause of this phenomenon is not the one-hot encoding \textit{per se}.
Rather, the problem is the combination of one-hot encoding and the use of axis-aligned decision rules based on one predictor.
One can, in fact, overcome the lack-of-support problem by utilizing decision rules based on linear combinations of the binary indicators created by one-hot encoding.
After all, for a categorical predictor $X$ and discrete set $\calC,$ the indicator $\ind{X \in \calC}$ is just the sum of binary indicators $\ind{X = c}$ for each $c \in \calC.$
More generally, one can imagine running BART using \textit{oblique} regression trees built with decision rules of the form $\{\phi^{\top}\bx_{\text{cont}} \in (-\infty, c)\}$ where $\bx_{\text{cont}}$ is the vector of continuous predictors.
\revised{
Whether BART ensembles built using oblique trees offer any practical or theoretical advantages over axis-aligned trees is a decidedly open question, which we leave to future work.}


\section*{Acknowledgements}
I would like to thank Prof.\ Cecilia Balocchi (Edinburgh) for several helpful discussions about flexible spatial modeling of areal crime data that motivated this work.
I am deeply grateful to my colleagues Profs.\ Keith Levin and Jun Zhu at Wisconsin for many productive conversations about networks, spanning trees, and spatial clustering.
Thanks are also due to Susan Glenn, Ajinkya Kokandakar, and Kehui Yao, who provided immensely helpful feedback on early versions of the \textbf{flexBART} package.
Finally, I am indebted to Prof.\ Ed George for providing constructive feedback on an early draft of this manuscript.

This research was performed using the computing resources and assistance of the UW--Madison Center For High Throughput Computing (CHTC) in the Department of Computer Sciences. 
The CHTC is supported by UW--Madison, the Advanced Computing Initiative, the Wisconsin Alumni Research Foundation, the Wisconsin Institutes for Discovery, and the National Science Foundation, and is an active member of the OSG Consortium, which is supported by the National Science Foundation and the U.S. Department of Energy’s Office of Science.

Support was also provided by the University of Wisconsin--Madison, Office of the Vice Chancellor for Research and Graduate Education with funding from the Wisconsin Alumni Research Foundation.

{
\singlespacing
\small
\bibliographystyle{apalike}
\bibliography{flex_bart}
}

\newpage
\appendix

\renewcommand{\thefigure}{\thesection\arabic{figure}}
\renewcommand{\thetable}{\thesection\arabic{table}}
\renewcommand{\theequation}{\thesection\arabic{equation}}

\setcounter{figure}{0}
\setcounter{equation}{0}

\begin{center}
{
\Large
\textbf{Supplementary Materials}
}
\end{center}

\appendix
\renewcommand{\thesubsection}{\Alph{section}\arabic{subsection}}
\renewcommand{\thefigure}{\thesection\arabic{figure}}
\renewcommand{\thetable}{\thesection\arabic{table}}
\renewcommand{\theequation}{\thesection\arabic{equation}}

\setcounter{figure}{0}
\setcounter{table}{0}
\section{Additional figures and tables}
\label{app:additional_figures}
\subsection{Network splitting strategies}

\begin{figure}[H]
\begin{subfigure}{0.245\textwidth}
\centering
\includegraphics[width = \textwidth]{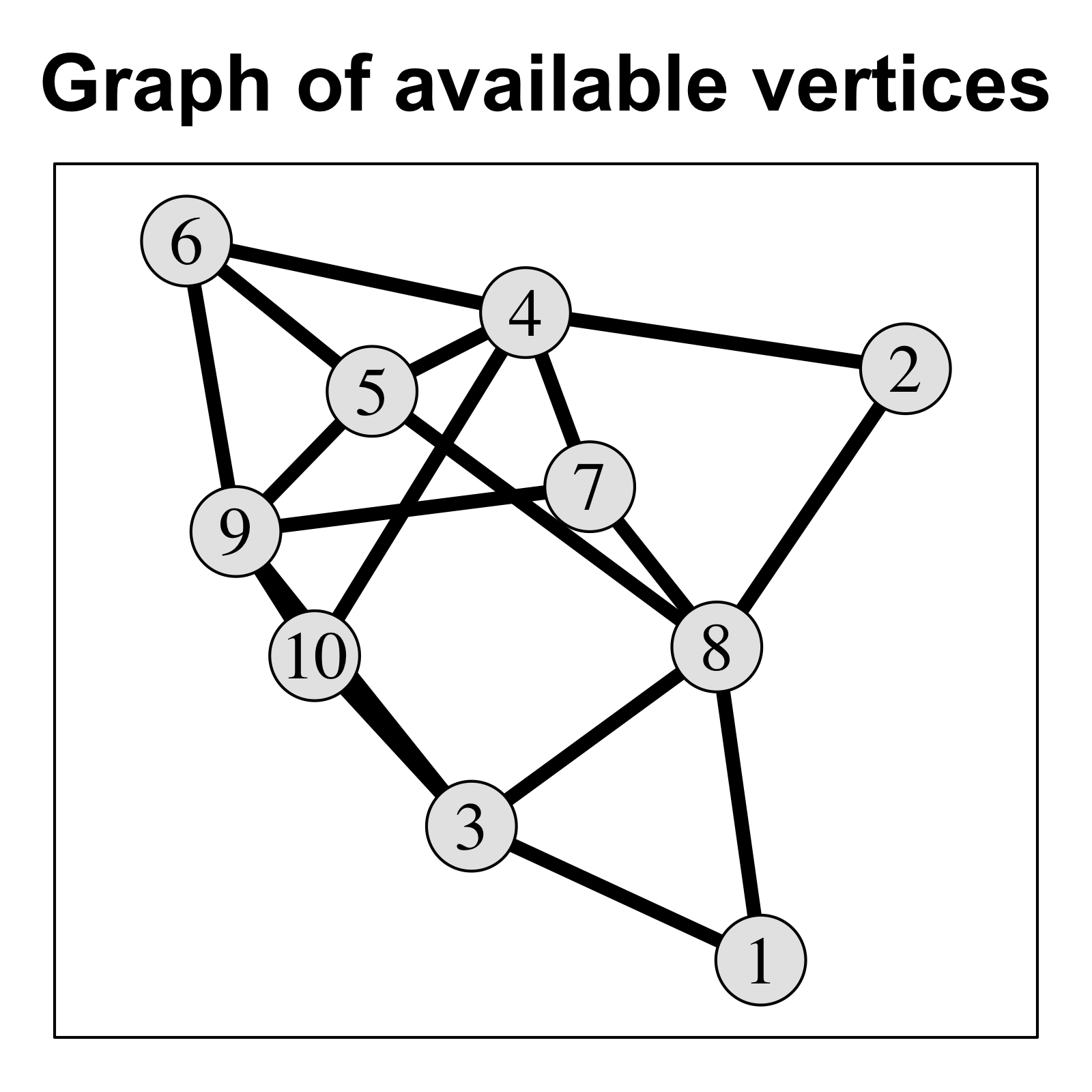}
\caption{}
\label{fig:mst_split_a}
\end{subfigure}
\begin{subfigure}{0.245\textwidth}
\centering
\includegraphics[width = \textwidth]{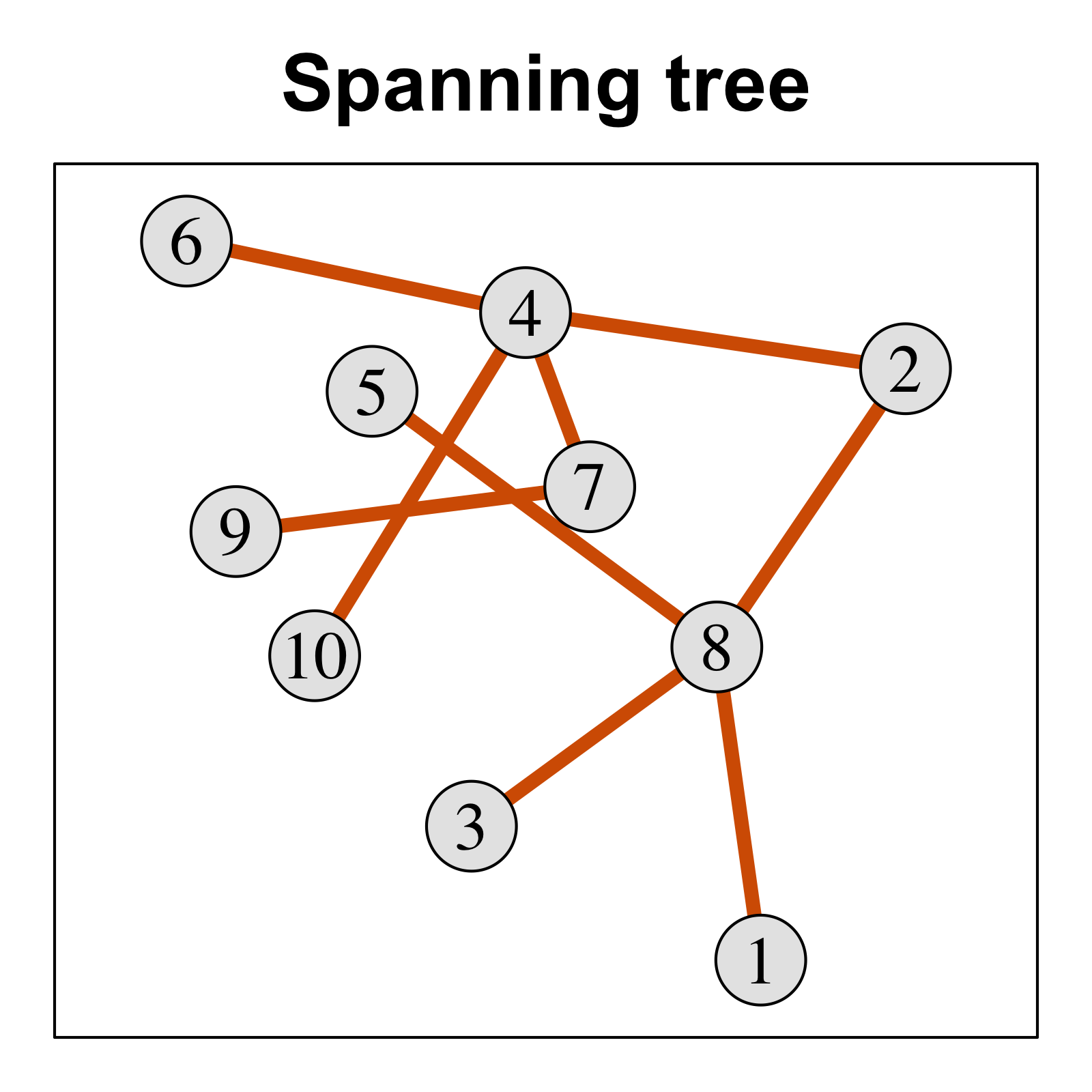}
\caption{}
\label{fig:mst_split_b}
\end{subfigure}
\begin{subfigure}{0.245\textwidth}
\centering
\includegraphics[width = \textwidth]{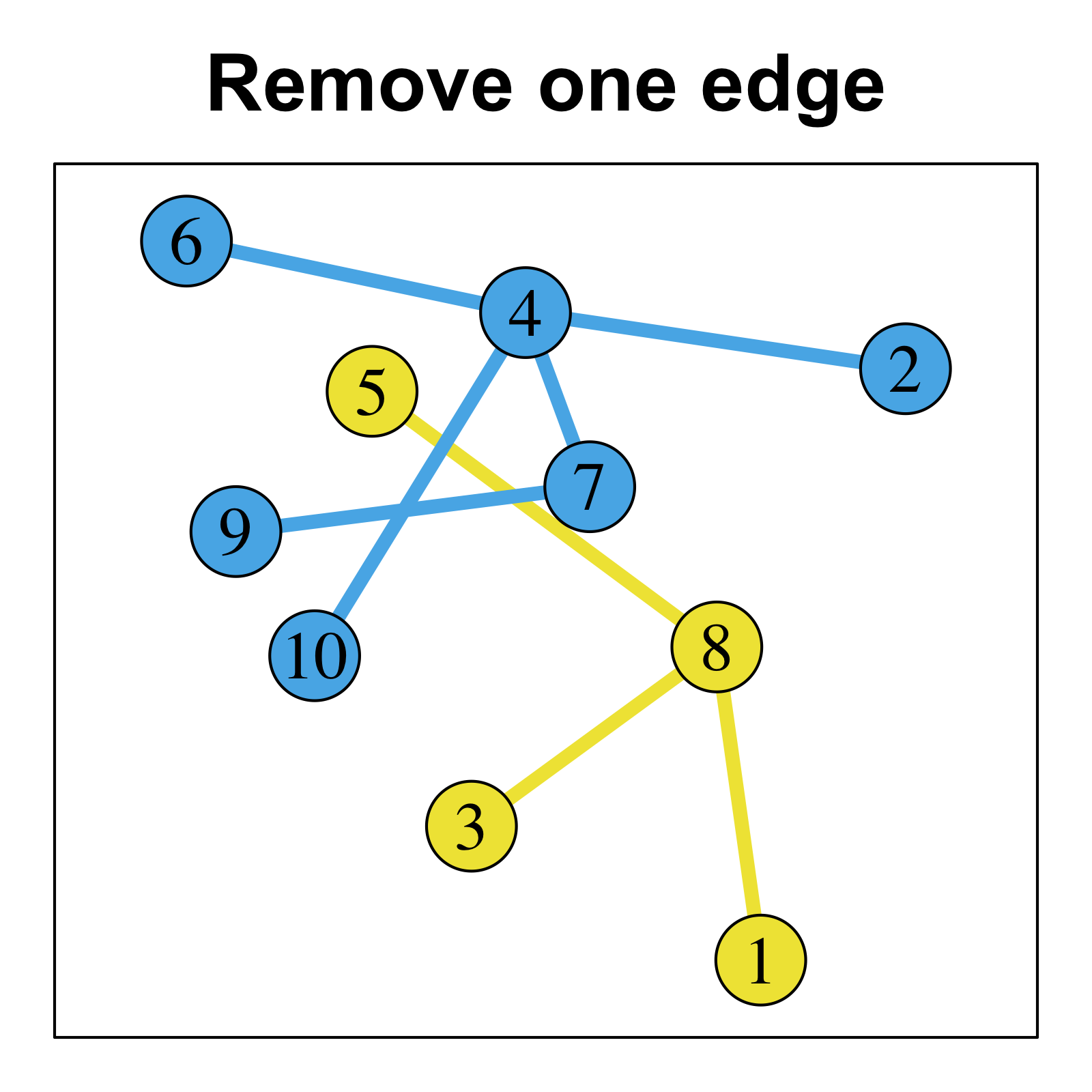}
\caption{}
\label{fig:mst_split_c}
\end{subfigure}
\begin{subfigure}{0.245\textwidth}
\centering
\includegraphics[width = \textwidth]{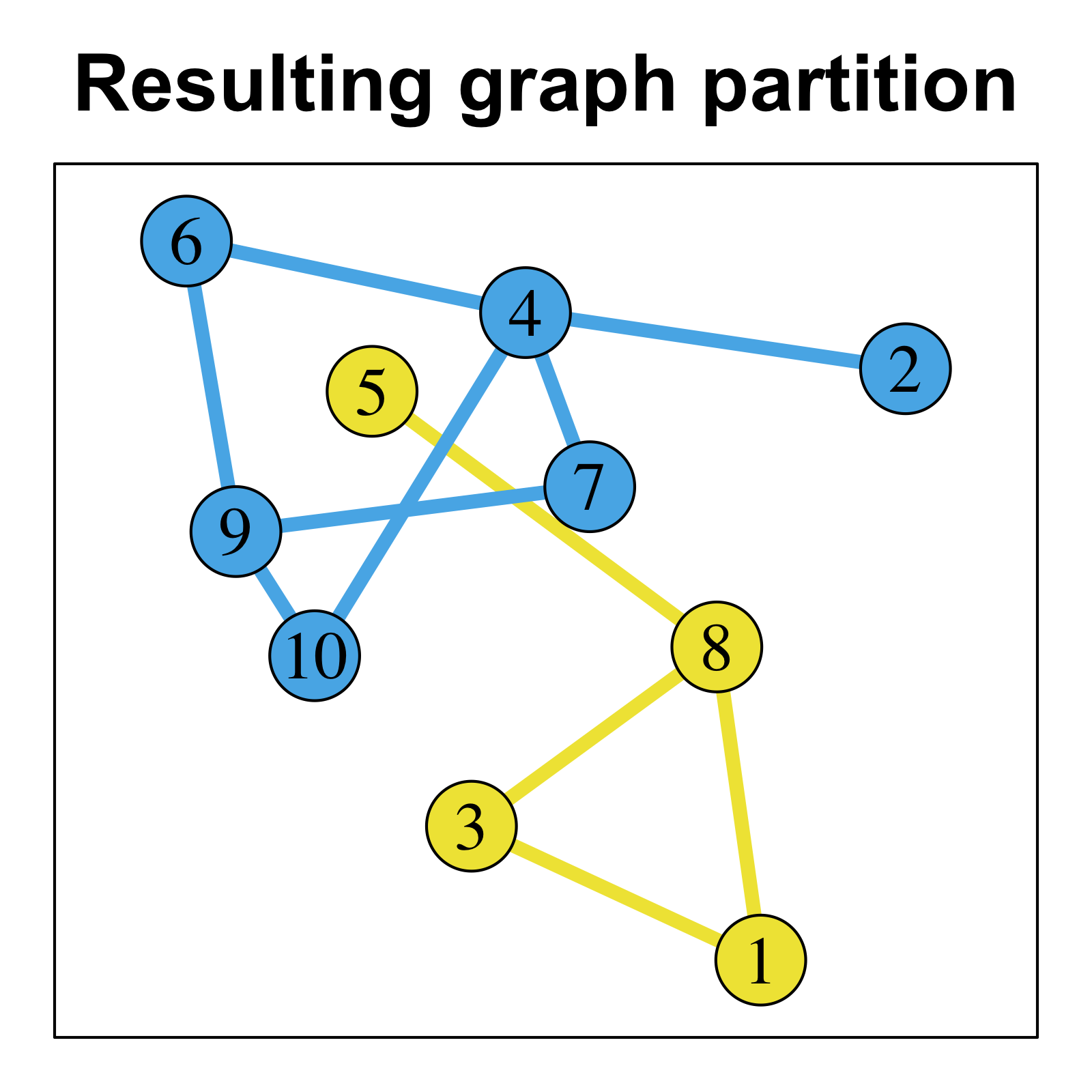}
\caption{}
\label{fig:mst_split_d}
\end{subfigure}
\caption{Cartoon illustration of one graph partitioning process. (a) Induced subgraph of levels available at a particular non-terminal node in the decision tree. (b) A spanning tree of the network. (c) Deleting one edge (in this case the edge $(2,8)$) from the spanning tree disconnects the tree and partitions the vertices into two subsets. (d) The subgraphs induced by the two vertex sets produced in (c) remain connected.}
\label{fig:mst_split}
\end{figure}

Our strategies \texttt{gs2} and \texttt{gs3} partition a network by drawing a spanning tree uniformly at random and then deleting an edge from that tree.
Figure~\ref{fig:mst_split} shows a cartoon illustration of this process.

\begin{figure}[ht]
\begin{subfigure}{0.32\textwidth}
\centering
\includegraphics[width = \textwidth]{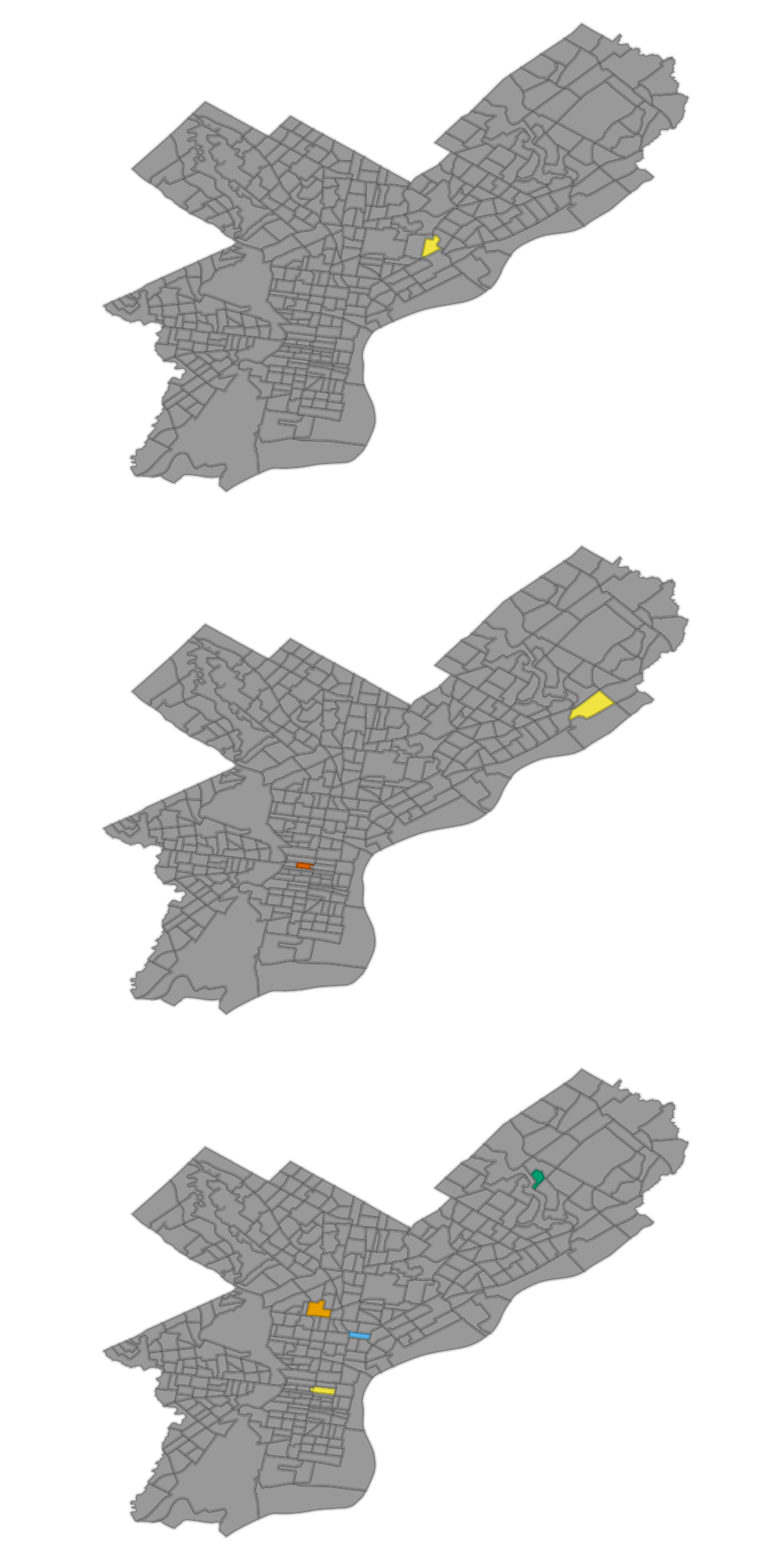}
\caption{}
\label{fig:philly_remove_one}
\end{subfigure}
\begin{subfigure}{0.32\textwidth}
\centering
\includegraphics[width = \textwidth]{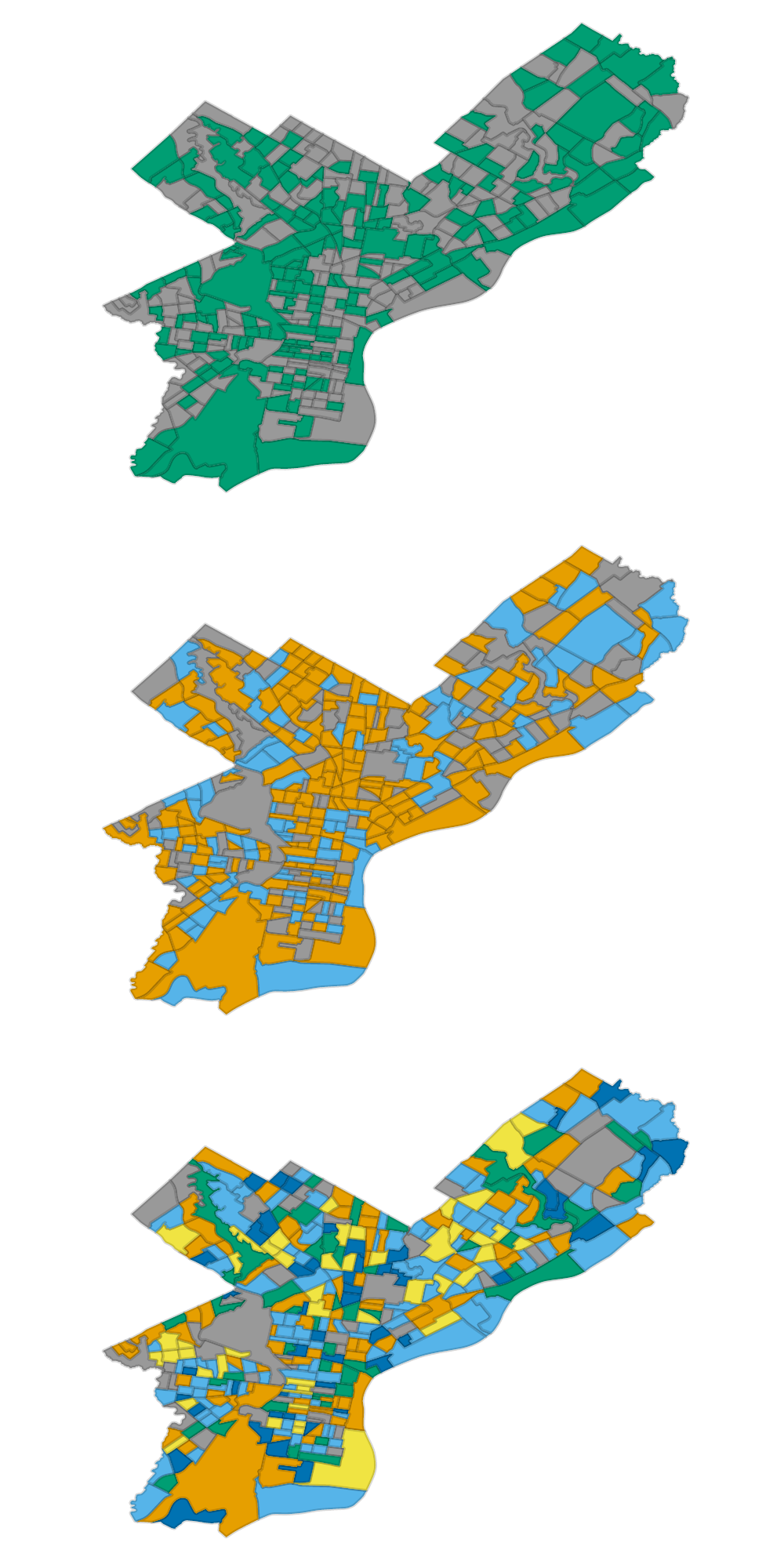}
\caption{}
\label{fig:philly_uniform_unordered}
\end{subfigure}
\begin{subfigure}{0.32\textwidth}
\centering
\includegraphics[width = \textwidth]{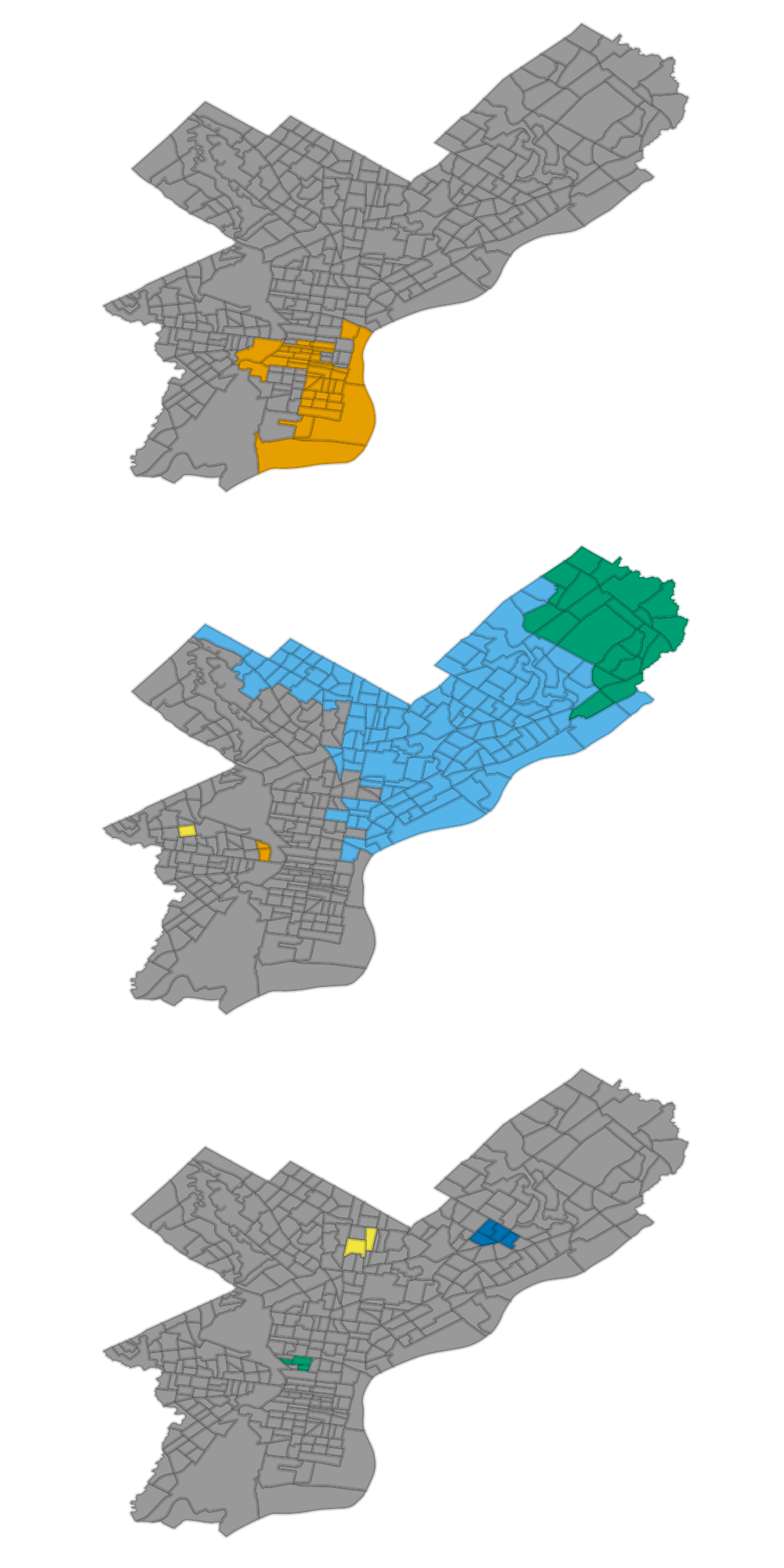}
\caption{}
\label{fig:philly_spanning_tree}
\end{subfigure}
\caption{Examples of partitions of the Philadelphia census tracts induced by regression trees drawn from (a) the BART prior after one-hot encoding; (b) without one-hot encoding and instead independently assigning levels to the left and right branches at random; and (c) using the network-splitting approach \texttt{gs2} described in Section 3.2 of the main text. We colored the census tracts according to their corresponding leaf in the regression tree.}
\label{fig:philly_partitions}
\end{figure}

Figure~\ref{fig:philly_partitions} illustrates typically partitions of the Philadelphia census tracts formed by the BAT prior obtained after one-hot encoding, independently assigning tracts to the left or right, and our network-splitting strategy \texttt{gs2}.
Although independently assigning tracts to the left or right branches allows the tree to pool data across multiple non-trivial subgroups of tracts, it clusters together geographically disparate tracts.
Our network-splitting strategies \texttt{gs1}--\texttt{gs4} force the tree to partition the tracts into spatially contiguous clusters.

\subsection{Synthetic data experiments}
\label{app:synthetic}

\begin{figure}[h!]
\centering
\begin{subfigure}{0.6\textwidth}
\centering
\includegraphics[width = \textwidth]{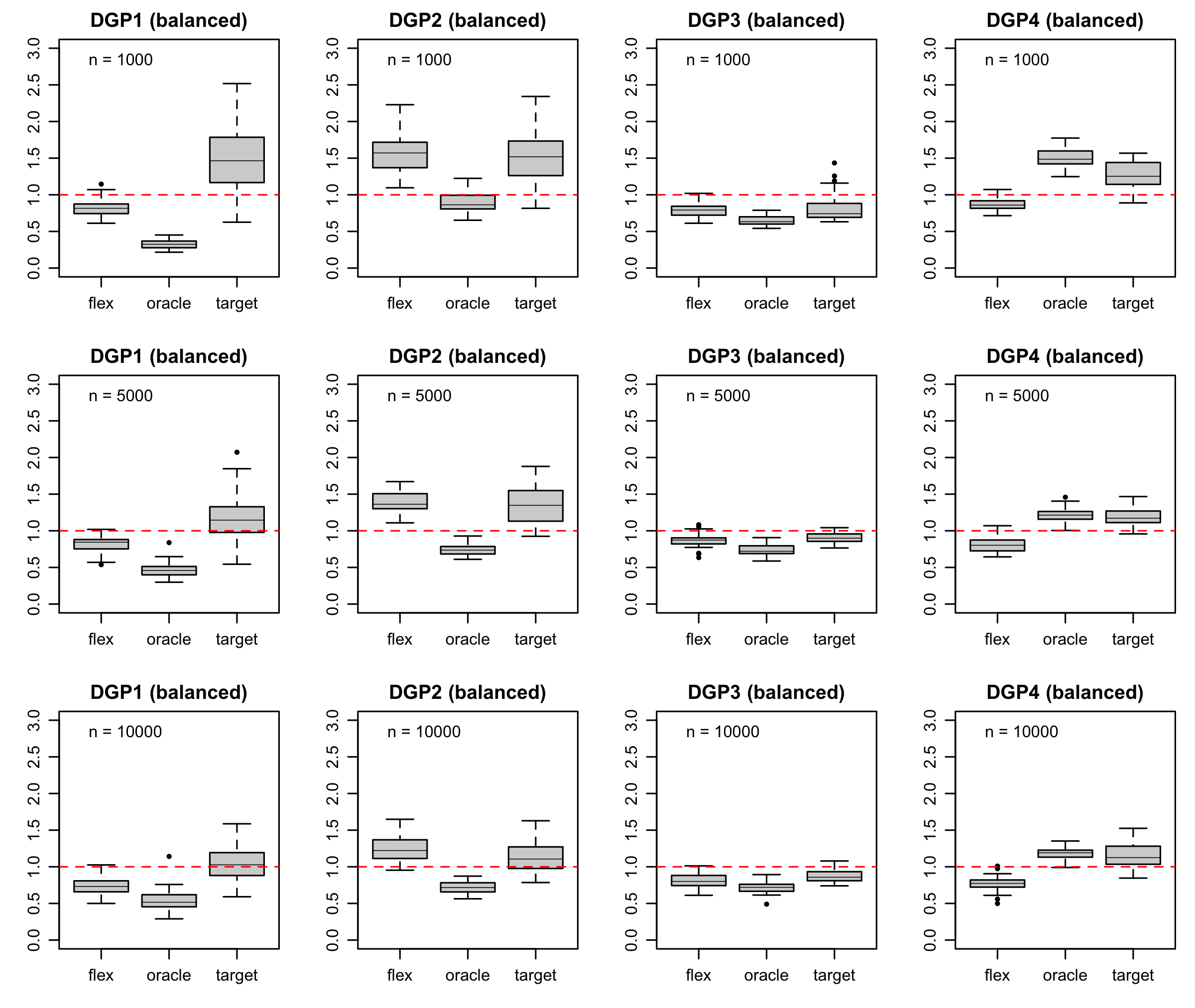}
\caption{}
\label{fig:grouping_advantage_unif}
\end{subfigure}

\begin{subfigure}{0.6\textwidth}
\centering
\includegraphics[width = \textwidth]{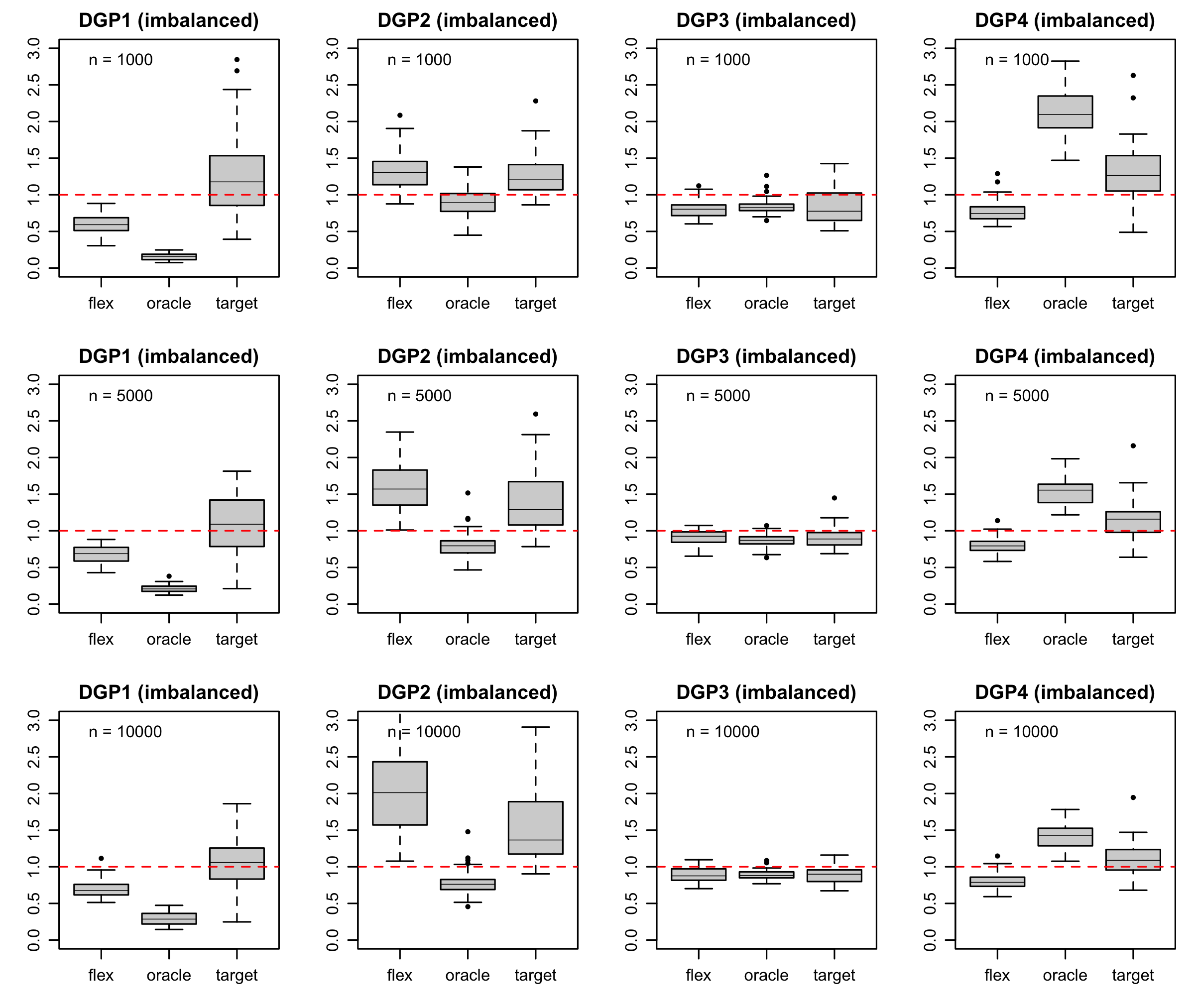}
\caption{}
\label{fig:grouping_advantage_imbal}
\end{subfigure}
\caption{Out-of-sample mean square errors relative to \texttt{BART} when the distribution of $X_{11}$'s levels are balanced (a) and imbalanced (b). Values less (resp.\ greater) than one indicate better (resp.\ worse) performance than \texttt{BART}.}
\label{fig:grouping_advantage}
\end{figure}

For each data generating process (DGP) $d \in \{1,2,3,4\}$ and $n \in \{1000, 5000, 10000\},$ we generated datasets of size $n$ from the model
\begin{align}
x_{1}, \ldots, x_{10} &\sim \unifdist{0}{1} & x_{11} &\sim \text{Multinomial}(p_{1}, \ldots, p_{10}) & y_{i} &\sim \normaldist{\mu_{d}(\bx_{i})}{1}.
\end{align}
We consider two different sets of class probability, a balanced setting in which each $p_{\ell} = 1/10$ and an imbalanced setting with
$$
(p_{1}, \ldots, p_{10}) = (0.01, 0.1, 0.02, 0.2, 0.15, 0.03, 0.05, 0.15, 0.25, 0.04).
$$
In the main text, we reported the results for the balanced setting with $n = 5000.$
Figures~\ref{fig:grouping_advantage_unif} and~\ref{fig:grouping_advantage_imbal} show the results for all $n$ and the two different settings of $(p_{1}, \ldots, p_{10}).$

For each DGP and pair of methods, we tested the null hypothesis that the two methods average out-of-sample error was identical using a two-sided paired t-test.
Table~\ref{tab:grouping_advantage} shows the p-values from these tests for the balanced setting with $n = 5000;$ the results for the imbalanced setting and other values of $n$ are similar.

\begin{table}
\centering
\caption{$p$-values from two-sided paired t-tests comparing each method in the synthetic data experiments. Bolded values indicate statistical significance at the 0.05 level after Bonferonni correction.}
\label{tab:grouping_advantage}
\scriptsize
\begin{tabular}{lcccc} \hline
\multicolumn{5}{c}{DGP1} \\ \hline
~ & flexBART & BART & \texttt{oracleBART} & \texttt{targetBART} \\ \hline
\texttt{flexBART} & -- & 5.5e-13$^{\star}$ & 6.5e-22$^{\star}$ & 1.7e-10$^{\star}$ \\
\texttt{BART} & -- & -- & 2.15e-38$^{\star}$ & 5.8e-4$^{\star}$ \\
\texttt{oracleBART} & -- & -- & -- & 3.9e-23$^{\star}$ \\ \hline
\multicolumn{5}{c}{DGP2} \\ \hline
~ & flexBART & BART & \texttt{oracleBART} & \texttt{targetBART} \\ \hline
\texttt{flexBART} & -- & 3e-27$^{\star}$ & 3.2e-36$^{\star}$& 0.043 \\
\texttt{BART} & -- & -- & 3.9e-25$^{\star}$ & 1.1e-13$^{\star}$ \\
\texttt{oracleBART} & -- & -- & -- & 6.6e-24$^{\star}$ \\ \hline
\multicolumn{5}{c}{DGP3} \\ \hline
~ & flexBART & BART & \texttt{oracleBART} & \texttt{targetBART} \\ \hline
\texttt{flexBART} & -- & 7.6e-14$^{\star}$ & 5.76e-12$^{\star}$ & 0.016 \\
\texttt{BART} & -- & -- & 5.4e-28$^{\star}$ & 5.8e-13$^{\star}$ \\
\texttt{oracleBART} & -- & -- & -- & 3.8e-23$^{\star}$ \\ \hline
\multicolumn{5}{c}{DGP4} \\ \hline
~ & flexBART & BART & \texttt{oracleBART} & \texttt{targetBART} \\ \hline
\texttt{flexBART} & -- & 2.4e-18$^{\star}$ & 1.3e-32$^{\star}$ & 6.4e-25$^{\star}$ \\
\texttt{BART} & -- & -- & 5.4e-28$^{\star}$ & 5.7e-13$^{\star}$ \\
\texttt{oracleBART} & -- & -- & -- & 3.8e-23$^{\star}$ \\ \hline
\end{tabular}
\end{table}

\subsection{Network-linked regression}
\label{sec:network_regression}

\begin{figure}[ht]
\centering
\begin{subfigure}{0.48\textwidth}
\centering
\includegraphics[width = 0.75\textwidth]{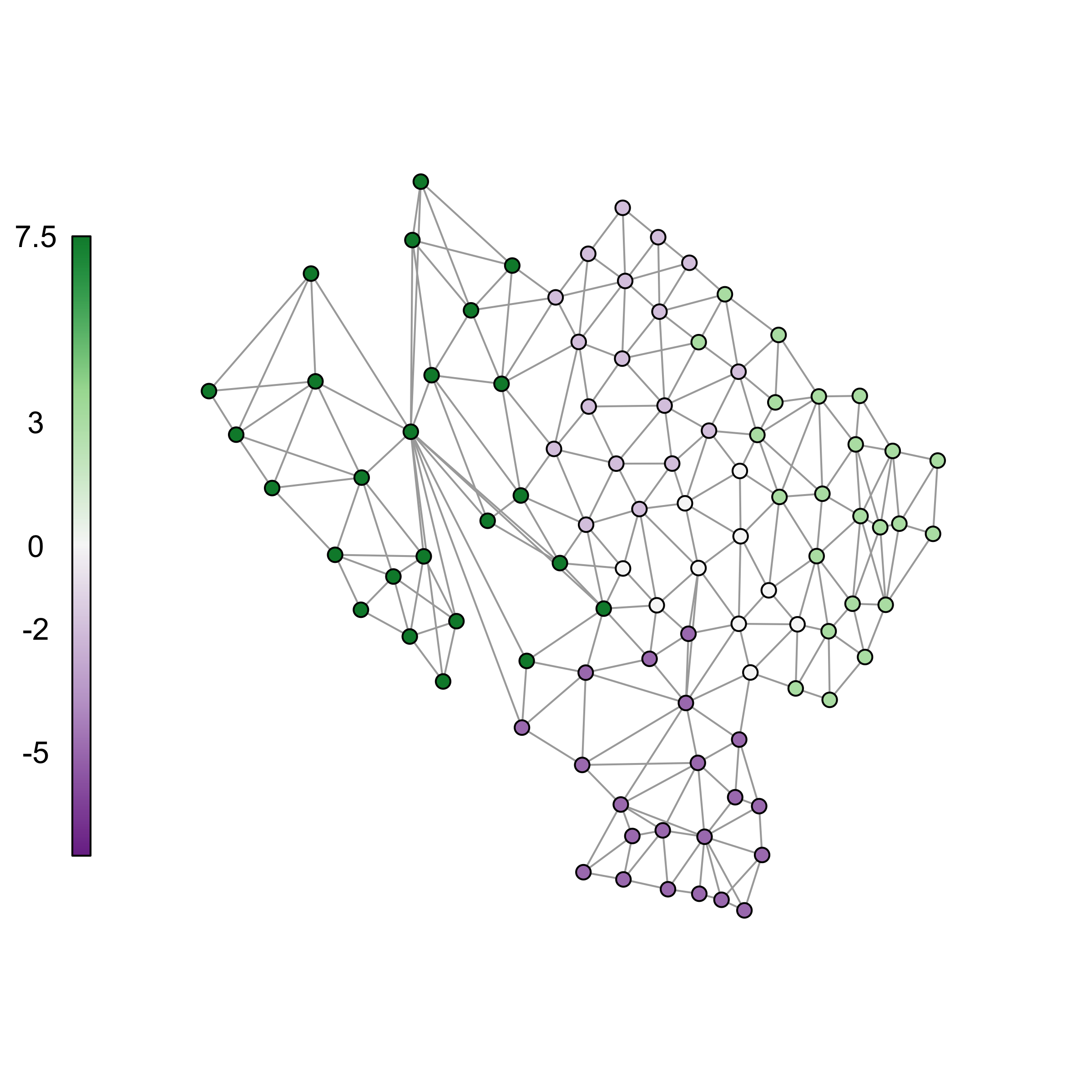}
\caption{}
\label{fig:network_constant1}
\end{subfigure}
\begin{subfigure}{0.48\textwidth}
\centering
\includegraphics[width = 0.75\textwidth]{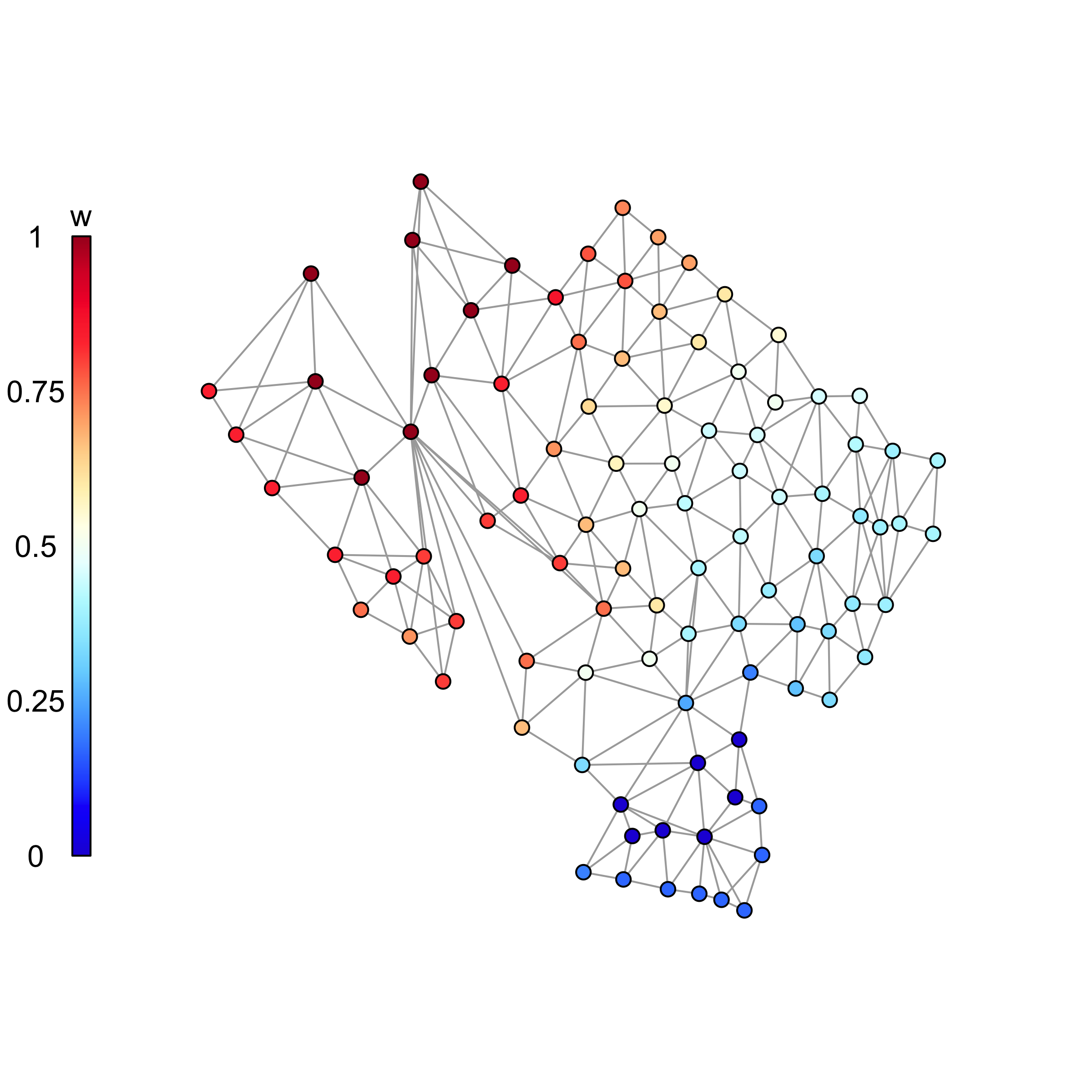}
\caption{}
\label{fig:network_fn}
\end{subfigure}
\caption{Piecewise constant function defined over a subset of the Philadelphia census tract network (a). Weights used to generate a smoothly varying function across the same network (b).}
\label{fig:network_dgp}
\end{figure}

For our network-linked regression experiments, we generated $t = 100$ noisy observations of a function defined over a subset of vertices in the Philadelphia census tract network.
For the first experiment, we used a piecewise constant function (Figure~\ref{fig:network_constant1}) and for the second experiment, we used a function that smoothly interpolated between two base functions.
Specifically, at vertex $v,$ we generated noisy evaluations of a function $g(\bx, v) = w_{v}g_{0}(\bx) + (1 - w_{v})g_{1}(\bx),$ where the vertex weight $w_{v}$ varied smoothly from 0 to 1 (Figure~\ref{fig:network_fn}) and 
\begin{align*}
g_{0}(\bx) &= 3x_{1} + (2 - 5 \times \ind{x_{2} > 0.5})\times\sin(\pi x_{1}) - 2\times \ind{x_{2} > 0.5}\\
g_{1}(\bx) &= 3 - 3\times \cos(6\pi x_{1}) \times x_{1}^{2} \times \ind{x_{1} > 0.6} - 10 \times \sqrt{x_{1}} \times \ind{x_{1} < 0.25}.
\end{align*}

For the piecewise constant experiment, we assessed each method's ability to evaluate the piecewise-constant at vertices seen during training and those held-out during training.
For the smoothly varying function, we assess each method's ability to evaluate the function at new $\bx$ values at vertices seen during training and at new $\bx$ values at the held-out vertices.
Table~\ref{tab:network_results} shows the root mean square error (RMSE) and predictive interval coverages for each method.

\begin{table}[H]
\centering
\caption{Average mean square error (MSE) and credible interval coverage (COV) across training and testing vertices in both network-linked regression experiments. For the smoothly varying function experiment, we report performance in estimating $g_{v}(\bx)$ at new $\bx$ values for vertices seen during training (Train) and vertices held-out of training (Test). Best performance is bolded and results are averaged across 50 simulation replications.}
\label{tab:network_results}
{
\begin{tabular}{lcccc|cccc} \hline
~ &\multicolumn{4}{c|}{Piecewise-constant} & \multicolumn{4}{c}{Smoothly varying} \\ 
~ &\multicolumn{2}{c}{Train} & \multicolumn{2}{c|}{Test} & \multicolumn{2}{c}{Train} & \multicolumn{2}{c}{Test} \\
Method & MSE& COV& MSE& COV& MSE & COV & MSE& COV \\ \hline
\texttt{BART} & 35.384 &  0.034 & 39.036 & 0.007 & 0.410 & 0.549 & 0.953 & 0.229 \\
\texttt{flexBART\_unif} & \textbf{0.010} & 0.945 & 29.791 & 0.516 & 0.311 & 0.765 & 1.866 & 0.524 \\
\texttt{flexBART\_gs1} & 0.086 & 0.913 & 6.995 & 0.769 & 0.199  & 0.750 & 0.226 & 0.736 \\
\texttt{flexBART\_gs2} & \textbf{0.010} & 0.945 & 3.708 & \textbf{0.793} & 0.144 & 0.842 & 0.204 & \textbf{0.835} \\ 
\texttt{flexBART\_gs3} & 0.009 & \textbf{0.952} & 4.471 & 0.784 & \textbf{0.137} & \textbf{0.848} & \textbf{0.191} & 0.824 \\ 
\texttt{flexBART\_gs4} & 0.008 & 0.949 & 4.384 & 0.782 & 0.141 & 0.836 & 0.204 & 0.802 \\ 
\texttt{BART\_ase1} & 0.012 & 0.926 & 46.459 & 0.200 & 0.597 & 0.588 & 0.977 & 0.497 \\
\texttt{BART\_ase3} & \textbf{0.010} & 0.942 & 4.281 & 0.556 & 0.173  & 0.818 & 0.248 & 0.778 \\
\texttt{BART\_ase5} & \textbf{0.010} & 0.947 & \textbf{3.678} & 0.631 & 0.169 & 0.818 & 0.258 & 0.808 \\ \hline
\end{tabular}
}
\end{table}

Tables~\ref{fig:network_constant1_pval} and~\ref{fig:network_fn_test2_pval} report p-values for comparing the average out-of-sample error in the two network-indexed regression experiments.

\begin{table}[h!]
\centering
\scriptsize
\caption{$p$-values from two-sided paired t-tests comparing each method in the piecewise constant network regression experiment. Starred values indicate statistical significance at the 0.05 level after Bonferonni correction.}
\label{fig:network_constant1_pval}
\begin{tabular}{lccccccccc}
~ & \texttt{BART} & \texttt{unif} & \texttt{gs1} & \texttt{gs2} & \texttt[{gs3} & \texttt{gs4} & \texttt{ase1} & \texttt{ase3} & \texttt{ase5} \\ \hline
\texttt{BART} & -- & 0.049  & 2.7e-9$^{*}$ & 1.5e-10$^{*}$ & 2.1e-10$^{*}$ & 1.7e-10$^{*}$ & 0.127 & 1.5e-10$^{*}$ & 1.1e-10$^{*}$ \\
\texttt{unif} & -- & -- & 1.3e-16$^{*}$ & 1.8e-20$^{*}$ & 1.3e-19$^{*}$ & 6.8e-20$^{*}$ & 1.9e-9$^{*}$ & 6.4e-21$^{*}$ & 1.4e-20$^{*}$ \\
\texttt{gs1} & -- & -- & -- & 2e-4$^{*}$ & 0.007 & 0.008 & 2.4e-23$^{*}$ & 0.011 & 5.9e-4$^{*}$ \\
\texttt{gs2} & -- & -- & -- & -- & 0.036 & 0.063 & 5.6e-27$^{*}$ & 0.208 & 0.933  \\ 
\texttt{gs3} & -- & -- & -- & -- & -- & 0.839 & 8.1e-26$^{*}$ & 0.715 & 0.034 \\
\texttt{gs4} & -- & -- & -- & -- & -- & -- & 3.8e-26$^{*}$ & 0.835 & 0.091 \\
\texttt{ase1} & -- & -- & -- & -- & -- & -- & -- &  1.6e-26$^{*}$ & 4.9e-27$^{*}$ \\
\texttt{ase3} & -- & -- & -- & -- & -- & -- & -- & -- & 0.163 \\ \hline
\end{tabular} 
\end{table}

\begin{table}[h!]
\centering
\scriptsize
\caption{$p$-values from two-sided paired t-tests comparing the out-of-sample mean square errors on test-set vertices in the smoothly-varying network regression experiment. Starred values indicate statistical significance at the 0.05 level after Bonferonni correction.}
\label{fig:network_fn_test2_pval}
\begin{tabular}{lccccccccc}
~ & \texttt{BART} & \texttt{unif} & \texttt{gs1} & \texttt{gs2} & \texttt{gs3} & \texttt{gs4} & \texttt{ase1} & \texttt{ase3} & \texttt{ase5} \\ \hline
\texttt{BART} & -- & 9.7e-19$^{*}$ & 3.5e-24$^{*}$ & 2.4-24$^{*}$ & 1.1e-24$^{*}$ & 1.1e-23$^{*}$ & 0.505 & 1.0e-22$^{*}$ & 1.4e-22$^{*}$ \\
\texttt{unif} & -- & -- & 5.6e-29$^{*}$ & 8.4e-30$^{*}$ & 1.1e-29$^{*}$ & 2.5e-29$^{*}$ & 3.9e-17$^{*}$ & 5.8e-29$^{*}$ & 3.3e-28$^{*}$\\
\texttt{gs1} & -- & -- & -- & 0.003 & 2.1e-05$^{*}$ & 0.021 & 2.7e-24$^{*}$ & 0.008 & 0.004 \\
\texttt{gs2} & -- & -- & -- & -- & 0.02  & 0.9.08 & 2.0e-24$^{*}$ & 1.5e-9$^{*}$ &1.0e-6$^{*}$  \\ 
\texttt{gs3} & -- & -- & -- & -- & -- & 0.059 & 4.6e-25$^{*}$ & 9.7e-10$^{*}$ & 5.5e-10$^{*}$ \\
\texttt{gs4} & -- & -- & -- & -- & -- & -- & 1.3e-23$^{*}$ & 1.3e-6$^{*}$ & 1.2e-5$^{*}$ \\
\texttt{ase1} & -- & -- & -- & -- & -- & -- & 4.9e-23$^{*}$ & 8.5e-23$^{*}$ \\
\texttt{ase3} & -- & -- & -- & -- & -- & -- & -- & 0.257 \\ \hline
\end{tabular} 
\end{table}

\subsection{Pitch framing}
\label{app:pitch_framing}

\begin{figure}[H]
\centering
\includegraphics[width = 0.7\textwidth]{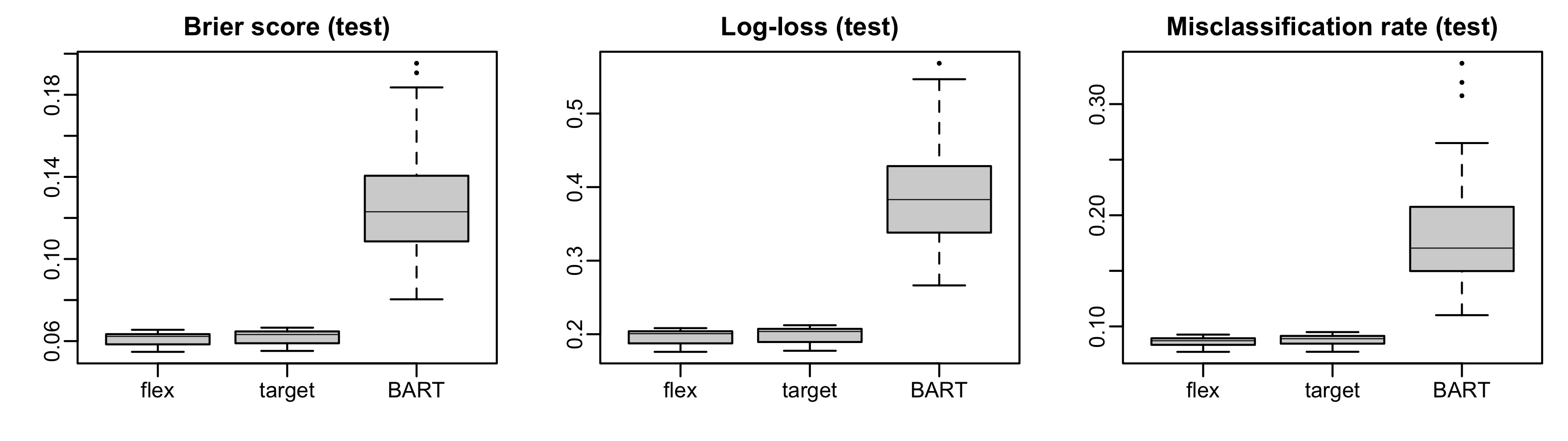}
\caption{Out-of-sample error across all seasons and cross-validation folds for the pitch framing data}
\label{fig:pitchFraming}
\end{figure}

Figure~\ref{fig:pitchFraming} compares the out-of-sample Brier score, log-loss, and misclassification rate for the four BART implementations on the pitch framing data.
Although \texttt{targetBART} appears to perform very similarly to \texttt{flexBART}, \texttt{flexBART} achieved smaller errors on every cross-validation fold.


\setcounter{equation}{0}
\section{Gibbs sampler derivation}
\label{app:calculations}
To describe the Gibbs sampler used to simulate posterior draws of the tree ensemble, we require some additional notation.
Recall that the decision tree $(T, \calD)$ partitions $\calX$ into disjoint, non-empty regions, one for each leaf in the tree $T.$
For each $\bx \in \calX,$ let $\ell(\bx)$ be the leaf corresponding to the region containing the point $\bx.$
For brevity, we will say that the leaf $\ell(\bx)$ contains the point $\bx.$
Further, by associating leaf $\ell$ with a scalar jump $\mu_{\ell},$ we can identify the regression tree $(T, \calD, \bmu)$ with a piecewise constant function.
Formally, we introduce the evaluation function $g: \calX \rightarrow \R,$ which is defined as
\begin{equation}
\label{eq:evaluation_function}
g(\bx; T, \calD, \bmu) = \mu_{\ell(\bx)},
\end{equation}
that returns the jump associated to the leaf containing the point $\bx.$

Given observations $(\bx_{1}, y_{1}), \ldots, (\bx_{n}, y_{n})$ and a decision tree $(T, \calD),$ let $I(\ell)$ be the set of indices for those observations contained in leaf $\ell.$ 
That is, $I(\ell) = \{i: \ell(\bx_{i}) = \ell\}.$
For compactness, we have suppressed the dependence of $\ell(\bx)$ and $I(\ell)$ on $(T, \calD)$ in our notation.
The collection of $I(\ell)$'s determines a mapping from the set of observations to the set of leafs in $T.$
This mapping plays a critical role in the conditional regression tree updates (see Equations~\eqref{eq:joint_factorization} and \eqref{eq:mu_posterior}).
When $n$ is large, computing the $I(\ell)$'s is the main computational bottleneck of each individual tree update.
As we describe later in Appendix~\ref{sec:redundant}, avoiding redundant re-computations of the $I(\ell)$'s yields considerable speed-up.

Recall that our model is $y \sim \normaldist{f(\bx)}{\sigma^{2}}$ and that we wish to estimate $f$ using a regression tree ensemble.
To this end, we formally introduce a collection $\calE = \{(T_{m}, \calD_{m}, \bmu_{m}): m = 1, \ldots, M\}$ of $M$ regression trees such that
$$
f(\bx) = \sum_{m = 1}^{M}{g(\bx; T_{m}, \calD_{m}, \bmu_{m})}.
$$
Let $\bm{f} = (f(\bx_{1}), \ldots, f(\bx_{n}))^{\top}$ be the vector of regression function evaluations at the observed inputs.
We specify independent and identical priors on each $(T_{m}, \calD_{m}, \bmu_{m})$ and use a Gibbs sampler to simulate posterior draws of $(\calE, \sigma) \vert \by.$
At a high-level, the sampler alternates between updating $\calE$ keeping $\sigma$ fixed and updating $\sigma$ keeping $\calE$ fixed.
When sampling $\calE,$ the sampler sequentially updates each regression tree conditionally fixing the other $M-1$ trees and $\sigma.$ 

\subsection{A Bayesian backfitting algorithm}
\label{sec:backfitting}

In what follows, we will describe the update for the $m$-th tree in the ensemble, conditionally given all other trees, the collection of which we denote as $\calE^{(-m)}.$
The conditional posterior distribution of the $m$-th tree has density
$$
p(T, \calD, \bmu \vert \by, \sigma, \calE^{(-m)}) \propto p(\by \vert T, \calD, \bmu, \sigma, \calE^{(-m)})p(T, \calD, \bmu).
$$
The sampler draws a tree from this distribution in two steps.
First, it draws a new decision tree $(T_{m}, \calD_{m})$ from the \textit{marginal} posterior distribution $(T, \calD) \vert \left[\by, \sigma, \calE^{(-m)}\right]$ with density
$$
p(T, \calD \vert \by, \sigma, \calE^{(-m)}) \propto \int{p(\by \vert T, \calD, \bmu, \sigma, \calE^{(-m)})p(T, \calD, \bmu)d\bmu}.
$$
Then, we draw new jumps from the conditional posterior of $\bmu \vert \left[T_{m}, \calD_{m}, \by, \sigma, \calE^{(-m)}\right].$

The conditional likelihood of the $m$-th tree plays a crucial role in both updates.
Given $\calE^{(-m)}$ and $\sigma,$ we have
\begin{equation}
\label{eq:conditional_tree_likelihood}
p(\by \vert T, \calD, \bmu, \sigma, \calE^{(-m)}) = (2\pi\sigma^{2})^{-n/2}\exp\left\{-\frac{1}{2\sigma^{2}}\sum_{i = 1}^{n}{\left[ r_{i} - g(\bx_{i}; T, \calD, \bmu)\right]^{2}}\right\},
\end{equation}
where
\begin{equation}
\label{eq:partial_residual}
r_{i} = y_{i} - \sum_{m' \neq m}{g(\bx_{i}; T_{m'}, \calD_{m'}, \bmu_{m'})}
\end{equation}
is the $i$-th partial residual.
Let $\bm{r} = (r_{1}, \ldots, r_{n})^{\top}$ denote the vector of partial residuals.

We can rewrite the sum over all observations in Equation~\eqref{eq:conditional_tree_likelihood} as the following double sum
\begin{equation}
\label{eq:leaf_loop_sum}
\sum_{i = 1}^{n}{\left[r_{i} - g(\bx_{i}; T, \calD, \bmu)\right]} = \sum_{\ell}{\sum_{i \in I(\ell)}{\left[r_{i} - \mu_{\ell}\right]}},
\end{equation}
where the outer sum is over the leafs of $T$ and the inner sum is over the observations contained in the leaf.
From this decomposition, we conclude that the conditional likelihood of the $m$-th regression tree factorizes over the leafs in $T:$
\begin{equation}
\label{eq:conditional_likelihood_factor}
p(\by \vert T, \calD, \bmu, \sigma, \calE^{(-m)}) = (2\pi\sigma^{2})^{-n/2} \times \prod_{\ell}{\exp\left\{-\frac{1}{2\sigma^{2}}\sum_{i \in I(\ell)}{\left[r_{i} - \mu_{\ell}\right]^{2}}\right\}}.
\end{equation}

Recall from Section 2.1 of the main text that we specified independent $\normaldist{\mu_{0}}{\tau^{2}}$ priors for each jump in $\bmu.$
We conclude, then, that the entire joint distribution of $(\by, T, \calD, \bmu)$ factorizes over the leaf nodes:
\begin{align}
\begin{split}
\label{eq:joint_factorization}
p(\by, T, \calD, \bmu \vert \sigma, \calE^{(-m)}) &= (2\pi\sigma^{2})^{-n/2} \times \prod_{\ell}{\left\{-\frac{1}{2\sigma^{2}}\sum_{i \in I(\ell)}{\left[r_{i} - \mu_{\ell}\right]^{2}}\right\}} \\
~ & ~\times \prod_{\ell}{\left[\left(2\pi\tau^{2}\right)^{-1/2}\exp\left\{-\frac{(\mu_{\ell} - \mu_{0})^{2}}{2\tau^{2}}\right\}\right]}.
\end{split}
\end{align}

The expression in Equation~\eqref{eq:joint_factorization} immediately implies that the jumps in $\bmu$ are conditionally independent given the decision tree $(T,\calD).$
In fact, we conclude that
$$
\mu_{\ell} \vert T, \calD, \by, \sigma, \calE^{(-m)} \sim \normaldist{P_{\ell}^{-1}\theta_{\ell}}{P_{\ell}^{-1}}
$$
where
\begin{align}
\label{eq:mu_posterior}
P_{\ell} &= \sigma^{-2}\lvert I(\ell)\rvert + \tau^{-2} & \theta_{\ell} &= \sigma^{-2}\sum_{i \in I(\ell)}{r_{i}} + \tau^{-2}\mu_{0}. 
\end{align}

Thus, conditional on the decision tree $(T, \calD),$ we must compute (i) the number of observations contained in the leaf and (ii) the sum of the partial residuals in each leaf in order to update the jumps.

\subsection{The grow/prune transition kernel}

Suppose that $(T_{m}, \calD_{m}) = (T, \calD)$ at the start of the decision tree update.
We draw a new value of the $m$-th decision tree with a Metropolis-Hastings step.
Specifically, we first propose a new tree $(T^{\star}, \calD^{\star})$ by randomly perturbing $(T,\calD).$
Then we accept the proposal and set $(T_{m}, \calD_{m}) = (T^{\star}, \calD^{\star})$ with probability
\begin{equation}
\label{eq:mh_prob}
\alpha((T, \calD) \rightarrow (T^{\star}, \calD^{\star})) = \min\left\{1, \frac{p(T^{\star}, \calD^{\star} \vert \by, \sigma, \calE^{(-m)})\revised{q(T, \calD \vert T^{\star}, \calD^{\star}, \by, \sigma, \calE^{(-m)})}}{p(T, \calD \vert \by, \sigma, \calE^{(-m)})\revised{q(T^{\star}, \calD^{\star} \vert T, \calD, \by, \sigma, \calE^{(-m)})}}\right\},
\end{equation}
where $q(\cdot \vert \cdot)$ is the transition density of the to-be-specific kernel.
If we reject the proposal, we leave the decision tree unchanged (i.e. we set $(T_{m}, \calD_{m}) = (T, \calD)).$

Before describing the transition kernel, observe that the ratio of posterior densities in Equation~\eqref{eq:mh_prob} decomposes as a product of a prior density ratio and a marginal likelihood ratio
$$
\frac{p(T^{\star}, \calD^{\star} \vert \by, \sigma, \calE^{(-m)})}{p(T, \calD \vert \by, \sigma, \calE^{(-m)})} =  \frac{p(T^{\star}, \calD^{\star})}{p(T, \calD)} \times \frac{p(\by \vert T^{\star}, \calD^{\star}, \sigma, \calE^{(-m)})}{p(\by \vert T, \calD, \sigma, \calE^{(-m)})}.
$$
In our simple regression setting, we obtain a closed-form expression for the marginal decision tree likelihood by integrating out $\bmu$ from Equation~\eqref{eq:joint_factorization}:
\begin{equation}
\label{eq:marginal_tree_likelihood}
p(\by \vert T, \calD, \sigma, \calE^{(-m)}) = \prod_{\ell}{\left[\tau^{-1} \times P(\ell)^{-1/2} \times \exp\left\{\frac{\Theta(\ell)^{2}}{2P(\ell)}\right\} \right]},
\end{equation}
where $P(\ell)$ and $\theta(\ell)$ are defined in Equation~\eqref{eq:mu_posterior}.

\begin{figure}[ht]
\centering
\includegraphics[width = 0.9\textwidth]{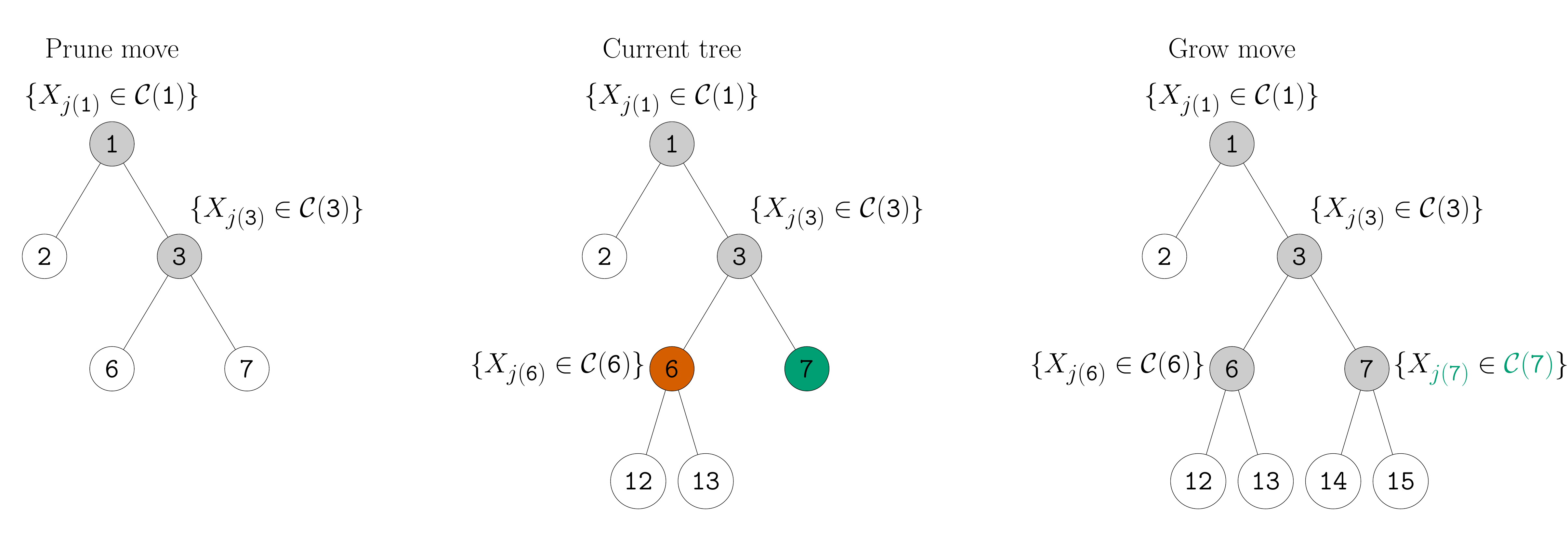}
\caption{Cartoon illustration of the grow and prune proposals used to draw new decision trees in our sampler. Observe that at most two leaf nodes are changed in both proposals. To implement the grow/prune kernel, one must be able to sample decision rules from the prior.}
\label{fig:grow_prune}
\end{figure}

We implemented a simple kernel that randomly grows or prunes a tree (see Figure~\ref{fig:grow_prune} for a cartoon illustration). 
Notice that we can reverse a grow transition with a prune transition and vice versa.
Conditional on growing the tree, we select a leaf node uniformly at random, attach two new children to the selected leaf, and draw a corresponding decision rule. 
Conditional on pruning the tree, we select a so-called ``no grandchild'' node uniformly at random, remove its two children from the tree, and delete its decision rule.

Growing and pruning change at most two leaves in the tree, leaving the rest unchanged.
In Figure~\ref{fig:grow_prune}, for instance, the current tree is otherwise identical to the pruned tree except at node $\texttt{6}.$
Similarly, the current tree differs from the grown tree only at node $\texttt{7}.$
The local nature of the transition kernel introduces considerable cancellation in the MH acceptance probability. 
We derive these acceptance probabilities for grow moves and prune moves below.

\textbf{Grow moves.} Suppose first that we grew $(T, \calD)$ from a leaf $\nx$ to obtain the proposal $(T^{\star}, \calD^{\star}).$
In the proposed tree, let $\{X_{j(\nx)} \in \calC(\nx)\}$ be the random decision rule associated with $\nx.$
To obtain $(T^{\star}, \calD^{\star})$ from $(T,\calD)$ from our kernel, we must (i) decide to grow, (ii) select $\nx$ uniformly at random; and (iii) draw the pair $(j(\nx), \calC(\nx))$ from \revised{a proposal distribution.
In our kernel, we do not allow the decision to grow or the choice of $\nx$ to depend on the fit provided by the other trees in the ensemble.
But we allow, for now, the possibility that the new decision rule is proposed in a way that depends on $\by, \calE^{(-m)}$ and $\sigma$ in addition to the current tree structure $(T, \calD).$}
The transition density for proposing $(T^{\star}, \calD^{\star})$ grown from $(T, \calD)$ is therefore
$$
q(T^{\star}, \calD^{\star} \vert T, \calD, \by, \sigma, \calE^{(-m)}) = q(\text{grow} \vert T) \times \texttt{n\_leaf}(T)^{-1} \times \revised{q(j(\nx), \calC(\nx) \vert T, \calD, \by, \sigma, \calE^{(-m)})}
$$
where $q(\text{grow} \vert T)$ is the probability of attempting a grow move; $\texttt{n\_leaf}(T)$ counts the number of leaf nodes in $T;$ and $\revised{q(j(\nx), \calC(\nx) \vert T, \calD, \by, \sigma, \calE^{(-m)})}$ is the proposal density for the new decision rule.
To compute the acceptance probability, we need the density of the reverse transition, which is
$$
q(T, \calD \vert T^{\star}, \calD^{\star}, \by, \sigma, \calE^{(-m)}) = q(\text{prune} \vert T^{\star}) \times \texttt{n\_nog}(T^{\star})^{-1},
$$
where $\texttt{n\_nog}(T^{\star})$ counts the number of non-terminal nodes in $T^{\star}$ whose children are leaf nodes (i.e., the nodes with ``\textbf{no} \textbf{g}randchildren'' in $T^{\star}$).

If $(T^{\star}, \calD^{\star})$ is obtained by growing $(T, \calD)$ from $\texttt{nx},$ the ratio of prior densities turns out to be
$$
\frac{p(T^{\star})p(\calD^{\star} \vert T^{\star})}{p(T)p(\calD \vert T)} = \frac{p(j(\texttt{\nx}), \calC(\texttt{nx})) \times \left[0.95(1 + d(\texttt{nx}))^{-2}\right] \times \left[1 - 0.95(2 + d(\texttt{nx}))^{-2}\right]^{2}}{\left[1 - 0.95(1 + d(\texttt{nx}))^{-2}\right]},
$$
where $d(\texttt{nx})$ is the depth of the node $\texttt{nx}$ in tree $T.$
Hence for grow moves, we compute
\begin{align}
\begin{split}
\label{eq:grow_prior_trans}
\frac{p(T^{\star},\calD^{\star})q(T, \calD \vert T^{\star}, \calD^{\star})}{p(T, \calD)q(T^{\star}, \calD^{\star} \vert T, \calD)} &= \frac{q(\text{prune} \vert T^{\star})/\texttt{n\_nog}(T^{\star})}{q(\text{grow} \vert T)/\texttt{n\_leaf}(T)} \times \frac{p(\texttt{j(\nx)}, \cutset(\nx))}{q(\texttt{j(\nx)}, \cutset(\nx) \vert T, \calD)} \\
~ & \times \frac{\left[0.95(1 + d(\texttt{nx}))^{-2}\right] \times \left[1 - 0.95(2 + d(\texttt{nx}))^{-2}\right]^{2}}{\left[1 - 0.95(1 + d(\texttt{nx}))^{-2}\right]}.
\end{split}
\end{align}

Turning to the ratio of marginal decision tree likelihoods, notice that there is substantial overlap in the set of leafs in $T$ and $T^{\star}.$
The only differences are (i) $\nx$ is a leaf in $T$ but not $T^{\star}$ and (ii) $T^{\star}$ contains two leaf nodes (the children of $\nx$ in $T^{\star},$ which we call $\nxl$ and $\nxr$) that are not themselves contained in $T.$
As a result, we have
\begin{align}
\begin{split}
\label{eq:grow_lil}
\frac{p(\by \vert T^{\star}, \calD^{\star}, \sigma, \calE^{(-m)})}{p(\by \vert T, \calD, \sigma, \calE^{(-m)})} &= \tau^{-1} \times \left(\frac{P(\nxr)P(\nxl)}{P(\nx)}\right)^{-\frac{1}{2}}  \\
~ & \times \exp\left\{\frac{\Theta(\nxl)^{2}}{2P(\nxl)} + \frac{\Theta(\nxr)^{2}}{2P(\nxr)} - \frac{\Theta(\nx)^{2}}{2P(\nxr)}\right\}
\end{split}
\end{align}

\revised{As noted in Section 2.2 of the main text, it is tempting to propose a new rule $\{X_{j(\nx)} \in \cutset(\nx)\}$ in an highly optimized way, for instance by deterministically proposing the optimal decision rule or by randomly selecting a rule from a narrow set of near-optimal candidates.
Doing so, however, renders the ratio $p(j(\nx), \cutset(\nx))/q(j(\nx), \cutset(\nx) \vert T, \calD, \by, \sigma, \calE^{(-m)})$ extremely small, which decreases the acceptance probability of grow moves, regardless of how much better the optimized $(T^{\star}, \calD^{\star})$ might fit the data than $(T, \calD).$
For that reason, we propose rules from the decision rule prior, which results in $p(j(\nx), \cutset(\nx))/q(j(\nx), \cutset(\nx) \vert T, \calD, \by, \sigma, \calE^{(-m)}) = 1.$}

\textbf{Prune moves}. Now suppose that we pruned $(T, \calD)$ by removing the children $\nxl$ and $\nxr$ of a no grandchild node of $\nx$ in $T$ to obtain $(T^{\star}, \calD^{\star}).$
Using essentially the same arguments for grow moves, we compute
\begin{align}
\begin{split}
\label{eq:prune_prior_trans}
\frac{p(T^{\star},\calD^{\star})q(T, \calD \vert T^{\star}, \calD^{\star})}{p(T, \calD)q(T^{\star}, \calD^{\star} \vert T, \calD)} &= \frac{q(\text{grow} \vert T^{\star})/\texttt{n\_leaf}(T^{\star})}{q(\text{prune} \vert T)/\texttt{n\_nog}(T)} \\
~ &\times \frac{\left[1 - 0.95(1 + d(\texttt{nx}))^{-2}\right]}{\left[0.95(1 + d(\texttt{nx}))^{-2}\right] \times \left[1 - 0.95(2 + d(\texttt{nx}))^{-2}\right]^{2}}.
\end{split}
\end{align}
and
\begin{align}
\begin{split}
\label{eq:prune_lil}
\frac{p(\by \vert T^{\star}, \calD^{\star}, \sigma, \calE^{(-m)})}{p(\by \vert T, \calD, \sigma, \calE^{(-m)})} &= \tau \times \left(\frac{P(\nx)}{P(\nxr)P(\nxl)}\right)^{-\frac{1}{2}} \\
&~\times\exp\left\{ \frac{\Theta(\nx)^{2}}{2P(\nx)} - \frac{\Theta(\nxl)^{2}}{2P(\nxl)} - \frac{\Theta(\nxr)^{2}}{2P(\nxr)}\right\}.
\end{split}
\end{align}
\revised{Note that Equation~\eqref{eq:prune_prior_trans} is based on proposing new decision rules in grow moves from the prior.}
Observe that to compute the MH acceptance ratio, we do not need to evaluate $P(\ell)$ and $\Theta(\ell)$ for every leaf in $T$ or $T^{\star}.$
To wit, we need only compute these quantities for exactly three nodes in $T^{\star}$ (resp.\ $T$) for grow (resp.\ prune) moves.
Crucially, this means that not every observation contributes to the MH acceptance ratio.
We exploit this fact in our implementation to avoid looping over all $n$ observations in our tree updates, which unnecessarily slows down the sampler.
In our implementation, we compute the log of the MH acceptance ratio calculations in the functions \texttt{grow\_tree} and \texttt{prune\_tree}.

\subsection{Mixing \& identifiability}
\label{sec:mixing_identifiability}

The individual trees in the BART ensembles are not identified, \revised{making it virtually impossible for the Gibbs sampler to mix over tree space.}
\revised{Practically, despite BART's failure to mix over tree space, we often obtain accurate point estimates and reasonably well-calibrated uncertainty intervals for evaluations of $f(\bx_{i})$ after running many short chains of the Gibbs sampler.}

\begin{figure}[ht]
\centering
\begin{subfigure}[b]{0.7\textwidth}
\centering
\includegraphics[width = \textwidth]{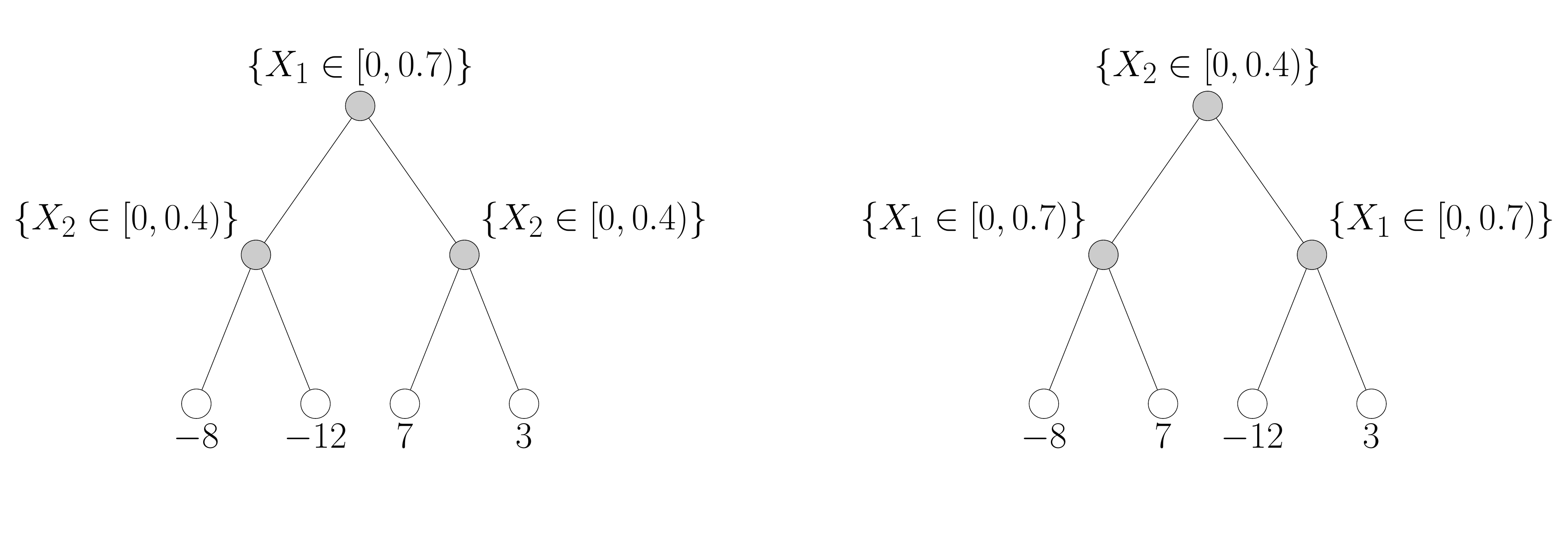}
\caption{}
\label{fig:overcomplete_trees}
\end{subfigure}
\\
\begin{subfigure}[b]{0.4\textwidth}
\centering
\includegraphics[width = \textwidth]{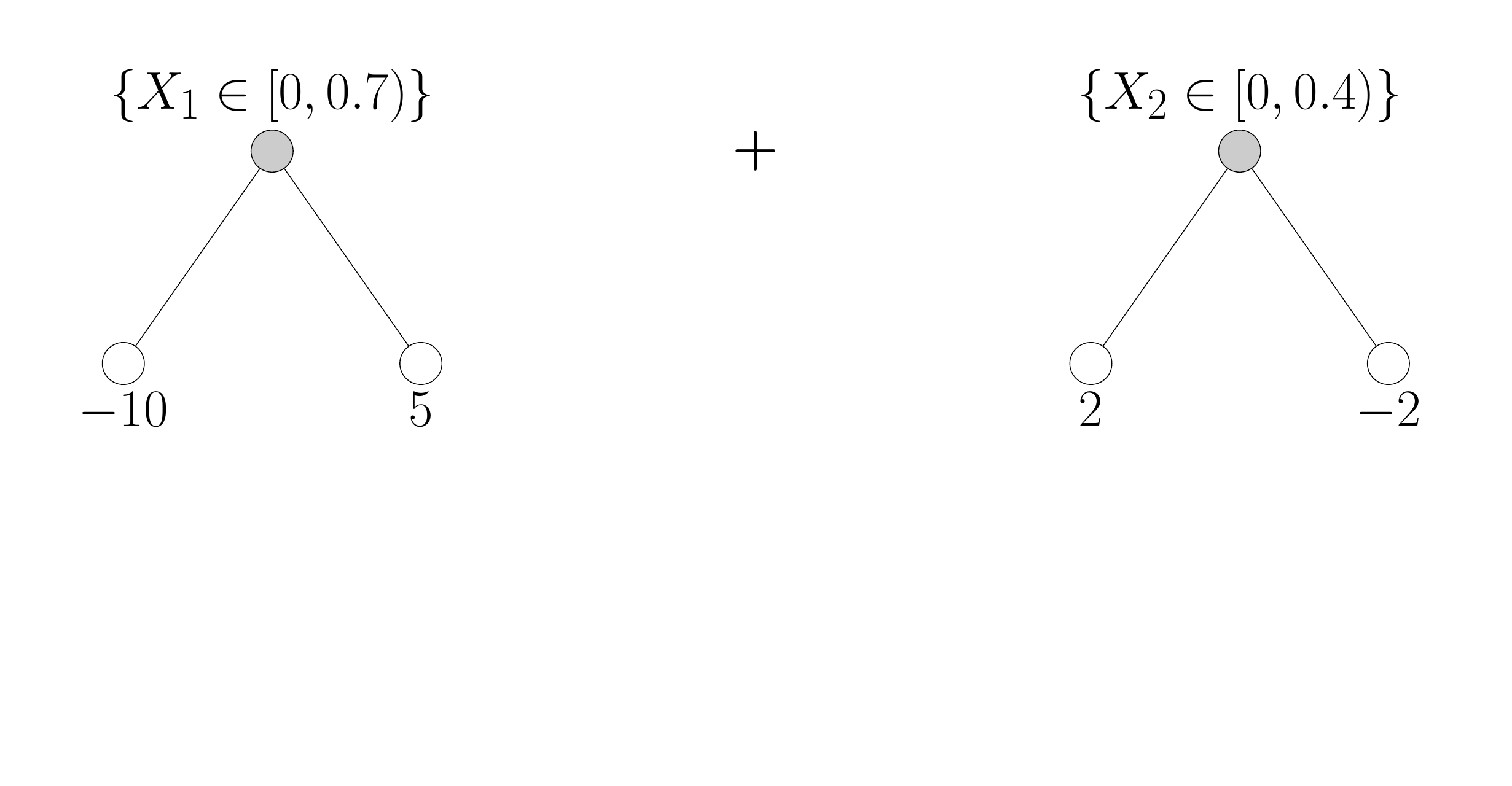}
\caption{}
\label{fig:overcomplete_sum1}
\end{subfigure}
\hspace{0.1\textwidth}
\begin{subfigure}[b]{0.4\textwidth}
\centering
\includegraphics[width = \textwidth]{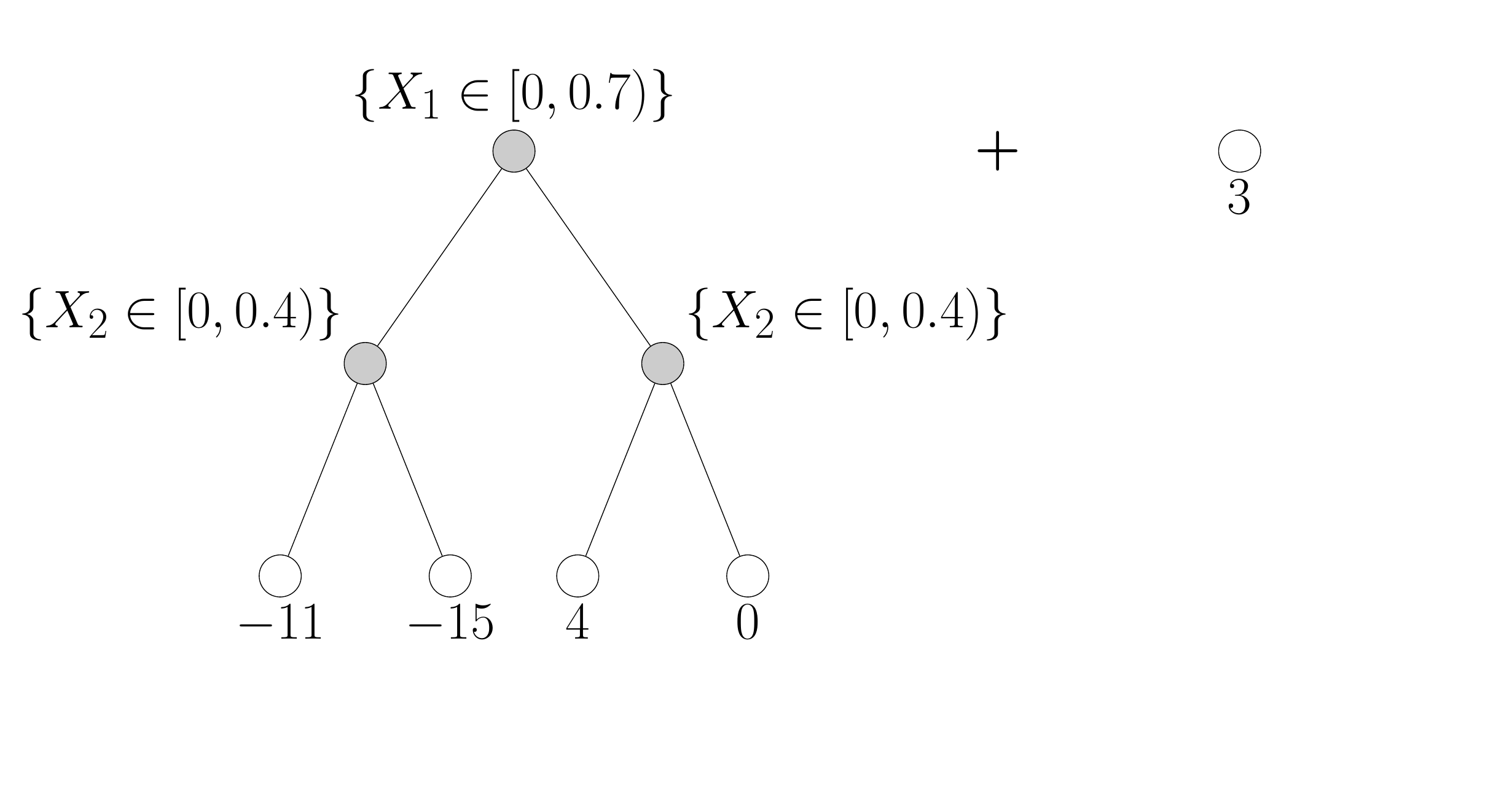}
\caption{}
\label{fig:overcomplete_sum2}
\end{subfigure}
\caption{(a) Piecewise constant functions can often be represented as a single regression tree in multiple ways. (b \& c) They can also be represented as sums of regression trees in multiple ways.}
\label{fig:overcomplete}
\end{figure}

\revised{To elaborate,} Figure~\ref{fig:overcomplete_trees} shows two representations of a step function defined over $[0,1]^{2}$ that assumes four different values..
\revised{Consider using a a simplified BART model with $m = 1$ trees to estimate this function.}
Insofar as the two trees in Figure~\ref{fig:overcomplete_trees} have the same structure and induce the same partition of the observations, they will have the same posterior densities.
To navigate between these two representations using only grow and prune moves, we must fully prune one tree back to the root node and then grow the other tree.
The intermediate states visited by the sampler will have rather low posterior density.
In other words, navigating between two modes requires traversing a valley of low probability, which is extremely difficult using only local moves.
Recently, \cite{Ronen2022_mixing} and \citet{KimRockova2023_mixing} have formalized this intuition, providing lower bounds of the mixing time for a single-tree model that are exponential in the number of observations.

%

The situation is even more fraught in the sum-of-trees setting.
For starters, the posterior landscape display a huge degree of symmetry.
Indeed, given a particular tree ensemble $\calE,$ we can obtain a notionally different ensemble by permuting the labels of the trees.
Beyond these symmetric representations, we can obtain equivalent fits to the data with very different tree ensembles.
For instance, in the context of Figure~\ref{fig:overcomplete_sum1}, we can define an two-tree ensemble in which the tree splitting on $X_{1}$ (resp.\ $X_{2})$ is the first (resp.\ second) tree and another in which it is the second (resp.\ first) tree.
And as Figure~\ref{fig:overcomplete_sum2} illustrates, we can represent the single trees in Figure~\ref{fig:overcomplete_trees} as a sum of similar tree and a stump, which consists only of the root node.
It is, in our opinion, essentially hopeless to expect our local grow/prune transition kernel to navigate efficiently between such representations.

The non-identifiability actually introduces more challenges beyond slow mixing.
It is exceedingly common to determine variable importance by counting the number of decision rules based on each predictor and to ``identify'' interactions by tracking the variables appearing in decision rules.
The fact that $\calE$ is not identified complicates the use of these heuristics.
Looking at the tree on the left of Figure~\ref{fig:overcomplete_trees}, we might conclude that (i) $X_{2}$ is more important than $X_{1}$ because it appears in two decision rules and that (ii) the function captures an interaction between $X_{1}$ and $X_{2}.$
Looking at the tree on the right, on the other hand, we would conclude that $X_{1}$ is more important than $X_{2}.$
The sum of trees representation in Figure~\ref{fig:overcomplete_sum1} paints yet another picture of the same function as a sum of two main effects, one for $X_{1}$ and one for $X_{2}.$

\setcounter{equation}{0}
\section{Implementation details}
\label{app:implementation_details}
In writing \textbf{flexBART}, we largely followed the example set by the authors of \textbf{BART}, writing \textbf{flexBART}'s core sampling functions in \textsf{C++} and interfacing those functions with \textsf{R} through \textbf{Rcpp} \citep{Rcpp}.
Although our overall design is similar to theirs, there are several key differences.
Briefly, we created a new, highly extensible class to represent decision rules and eliminated many redundant calculations performed by \textbf{BART}, which led to substantial run-time improvements.
Below, we outline some of the important class and data structures that we introduced to represent regression trees, decision rules, and the sets $I(\ell).$
Then, we describe how carefully tracking and updating our representations of the $I(\ell)$'s facilitated considerable speed-up relative to \textbf{BART}.

\subsection{Class \& data structures}
\label{sec:class_structure}

As described in \citet{Pratola2014_parallel}, \textbf{BART} represents regression trees as a collection of nodes, which are instances of a custom class containing only six data members: (i) pointers to separate class instances for the parent and children of the node; (ii) an integer for the decision variable index $j$; (iii) an integer recording the index of the cutpoint $c$\footnote{\textbf{BART} pre-computes a finite grid of cutpoints for each predictor instead of drawing them from $[0,1]$.}; and (iv) a double for the jump.
Armed with this lightweight class, \textbf{BART}'s authors wrote a recursive functions that traces the decision-following path of each $\bx \in \calX.$

Although the class and this function facilitate extremely fast and memory-efficient MCMC simulations, they are insufficient for representing regression trees like the ones described in Section 2 of the main text.
We consequently re-wrote the regression tree node class and replaced the two integers parametrizing the decision rule with an instance of a new class representing decision rules.
To accommodate our new classes, we rewrote the function that determines the leaf node corresponding to a give $\bx.$
Briefly, given an instance of the rule class and $\bx,$ that function checks whether the decision variable is continuous or categorical.
If the variable is continuous, the function checks whether the corresponding entry in $\bx$ is less than the cutpoint.
Otherwise, for a categorical splitting variable, the function checks whether the relevant entry in $\bx$ is assigned to the left child of the node.
We also introduced a separate function (\texttt{draw\_rule}) to draw new decision rules from the prior for use in the Metropolis-Hastings proposal.

\textbf{The \texttt{tree} class}. 
Regression tree nodes in our implementation are instances of a custom \textsf{C++} class, which we named $\texttt{tree}.$
The class data members include (i) pointers to the parent, left child, and right child of the node; (ii) an instance of a decision rule class (described below); and (iii) a double for the jump $\mu.$
Most of the class's function members are convenience functions for getting and setting individual data members.
For instance, the function \texttt{tree::set\_mu} assigns the value of the jump $\mu$ associated with the node while the function \texttt{tree::get\_nid} returns a canonical label for each node.
These labels are computed recursively: starting from the root node, which is labelled \texttt{1}, the left (resp.\ right) child of a node with label $\nx$ is $2\nx$ (resp.\ $2\nx + 1$).
The class also contains functions for computing summary statistics about the tree including the numbers of leaf nodes (\texttt{tree::get\_nbots}) and ``no grandchild'' nodes, whose children are leaf nodes (\texttt{tree::get\_nnog}). 

Beyond these basic functions, the class contains some more substantive functions.
There are two functions (\texttt{tree::birth} and \texttt{tree::death}) that perform the necessary book-keping associated with our grow/prune transition kernel.
For instance, when we accept a grow transition that adds two children to the node \nx, \texttt{tree::birth} creates two new class instances, one for each child of \nx; sets the parent pointers in each of these instances; and updates the left and right child pointers in \nx's \texttt{tree} instance.
Similarly, when we accept a prune transition that removes two nodes, \texttt{tree::prune} deletes the class instances for these nodes; and sets the child pointers of the parent to the null pointer.
The class additionally contains functions that read in a predictor vector $\bx$\footnote{Technically, the function takes two pointers as input, one for the continuous predictors and one for the categorical predictors} and return a pointer to the leaf node $\ell(\nx)$ containing $\nx$ (\texttt{tree::get\_bn}) and the value $g(\bx; T, \calD, \bmu)$ (\texttt{tree::evaluate}).
Both functions trace the full decision-following path from the root to the leaf $\ell(\bx).$
Although we do not need to call these in our main MCMC loop, we do utilize them when drawing posterior predictive samples of regression function evaluations $f(\bx^{\star})$ at new predictors $\bx^{\star}$.
Finally, the \texttt{tree} class contains two functions used to compute the set $\calA$ of available predictor values at each regression tree node, one for continuous predictors and one for categorical predictors.
Both functions work by recursing up the tree (by following parent pointers) and checking whether any ancestors of a given node also split on the same variable.
These functions are called when drawing a new decision rule for grow transitions.

In our implementation, we represent the regression tree ensemble $\calE$ with a vector of \texttt{tree} instances.

\textbf{The \texttt{rule\_t} class}. Unlike \textbf{BART}'s source code, which hardcoded decision rule information into the tree class, we created separate class called \texttt{rule\_t} encoding the decision rules.
Our class contains several data members including (i) the decision variable index $j$; (ii) a Boolean encoding whether that splitting variable is continuous or categorical; (iii) a cutpoint $c$; and (iv) two sets, \texttt{l\_vals} and \texttt{r\_vals}, that record the categorical levels assigned to the left and right child.
Note that the cutpoint is set and accessed only when $X_{j}$ is continuous and the two sets are populated only when $X_{j}$ is categorical.
Otherwise, they remain at default values ($c$'s default value is 0 and \texttt{l\_vals} and \texttt{r\_vals} default to \texttt{std::set<int>}'s of length zero). 

When we propose grow moves in our sampler, we draw a new decision rule using the function \texttt{draw\_rule}.
That function first selects the splitting variable and then checks whether it is continuous or categorical.
If it is continuous, the function calls the function \texttt{tree::get\_rg\_cont} to determine the interval of available values and then draws a cutpoint uniformly from that interval.
If the splitting variable is categorical, it uses \texttt{tree::get\_rg\_cat} to determine the set of available levels.
Then, \texttt{draw\_rule} checks whether the levels are network-structured.
If they are not, it repeatedly sweeps over the levels, assigning each to \texttt{l\_vals} with probability 0.5, until both \texttt{l\_vals} and \texttt{r\_vals} contain at least one element.
If the levels are network-structured, \texttt{draw\_rule} forms the induced subgraph $\calG[\calA];$ samples a uniform spanning tree with Wilson's algorithm; deletes an edge from the spanning tree; and assigns the vertices of the two connected components to \texttt{l\_vals} and \texttt{r\_vals}. 

\textbf{The \texttt{suff\_stat\_map} structure.} Recall from Equation~\eqref{eq:joint_factorization} that the joint density of the data \& $m$-th regression tree, conditional on all other regression trees and $\sigma,$ factorizes over the leaf nodes.
Consequently, all calculation required to update an individual regression tree involve determining which observations are contained in which leafs of a tree.
In a sense, the map from observations to leaf node is a sufficient statistic, conditional on $\sigma$ and $\calE^{(-m)}.$
To keep track of these maps, we created a separate structure, which we called \texttt{suff\_stat\_map}.
The structure is implemented as a \texttt{std::map} whose keys are integers corresponding to the canonical labels of the leaf nodes and whose values are integer vectors of observation indices.
In our implementation, we keep track of a vector of \texttt{suff\_stat\_map}'s, one for each regression tree.
During our MCMC simulation, we update both the regression trees and the sets $I(\ell)$ whenever we accept a grow or prune transition.
We wrote two function, \texttt{compute\_suff\_stat\_grow} and \texttt{compute\_suff\_stat\_prune} that compute the $I(\ell)$'s for the proposal tree $T^{\star}.$

\subsection{Avoiding redundant computations}
\label{sec:redundant}

In each iteration of our MCMC simulation, we must compute the vector of partial residuals $\br$ before updating the $m$-th tree.
Perhaps the simplest approach, based solely on Equation~\eqref{eq:partial_residual}, is to compute the fit of each tree at each observed predictor vector $\bx.$
That is, we would loop over the trees in $\calE^{(-m)}$ and all $n$ observations to compute each $g(\bx_{i}; T_{m'}, \calD_{m'}, \bmu_{m'})$ using the function \texttt{tree::evaluate}.
Since the trees tend to be relative shallow (thanks to the regularization prior), individual calls of \texttt{tree::evaluate} are quite fast.
Nevertheless, computing $M \times (M-1)\times n$ tree fits from scratch in every MCMC iteration gets expensive, especially when $n$ is large.
Instead, following \textbf{BART}'s example, we keep track of two quantities, \texttt{allfit\_train}, which is a running estimate of $\bm{f},$ and \texttt{residual}, which stores the \textit{full} residual $\by - \bm{f}.$
At the beginning of the $m$-th regression tree update, we temporarily remove the fit of the $m$-th tree from \texttt{allfit\_train} and add the tree fit to \texttt{residual}.
We then pass pointers to these arrays to other functions that perform the MH update.
Finally, after updating the regression tree, we compute the fit of the $m$-th tree, restore it to \texttt{allfit\_train}, and remove it from \texttt{residual}.
In this way, when updating the $m$-th regression tree, we need only evaluate the fit of that tree at every observation.

Although carefully updating running estimates of $\bm{f}$ and $\br$ avoids looping over all trees and all observations in every tree update, it still requires a loop over the observations to compute $g(\bx_{i}; T_{m}, \calD_{m}, \bmu_{m}).$
When $n$ is large, even this single loop can be prohibitively expensive.
It turns out that we can avoid calling \texttt{tree::evaluate} altogether by using our \texttt{suff\_stat\_map} structures.
Specifically, we can update the entries in \texttt{allfit\_train} and \texttt{residual} with an outer loop over each leaf and an inner loop over observations contained in the leaf. 
Within the outer loop, we use the class function \texttt{tree::get\_mu} to look up the value of the jump $\mu_{\ell}$ associated with the current leaf.
Then, within the same outer loop, we loop over the observations contained in the leaf and add or subtract (resp.\ add) $\mu_{\ell}$ to the entries of \texttt{allfit\_train} (resp.\ \texttt{residual}).
We perform a similar loop after the tree has been updated, except this time we subtract (resp.\ add) the jumps to the entries of \texttt{allfit\_train} (resp.\ \texttt{residual}).
While these loops introduce some computational overhead due to the non-contiguous access to \texttt{allfit\_train} and \texttt{residual}, they importantly avoid having to compute each $g(\bx_{i};T, \calD, \bmu)$ anew each iteration.
The speedup of \textbf{flexBART} relative to \textbf{BART} largely stems from this avoidance.

\end{document}